\documentclass[journal,comsoc]{IEEEtran}

\usepackage[T1]{fontenc}
\usepackage{cite}
\usepackage{amsmath,amssymb,amsfonts}
\usepackage{algorithmic}
\usepackage{graphicx}
\usepackage{textcomp}
\usepackage{xcolor}
\usepackage[font=small]{caption}
\usepackage{subcaption}
\usepackage{array}%

\ifCLASSINFOpdf
\else
\fi
\usepackage{amsmath}
\usepackage[cmintegrals]{newtxmath}
\newcommand{\figintroJSCC}{
  \begin{figure}[t]
    \centering
    \includegraphics[width=\columnwidth]{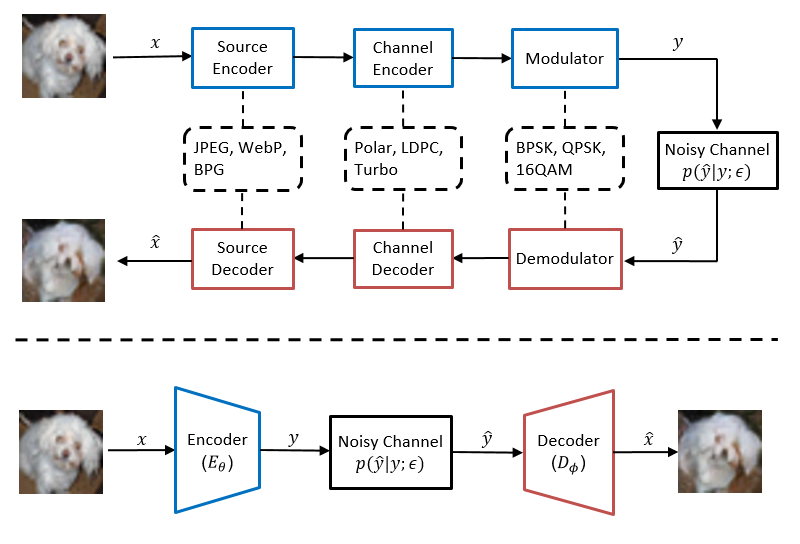}
    \caption{Block diagrams of wireless image transmission schemes. Top: Traditional separate source and channel coding scheme. Bottom: Deep learning based joint source channel coding (JSCC) scheme.}
    \label{fig:fig_intro}
  \end{figure}
}

\newcommand{\figofdmlarge}{
  \begin{figure*}[t]
    \centering
    \includegraphics[width=\linewidth]{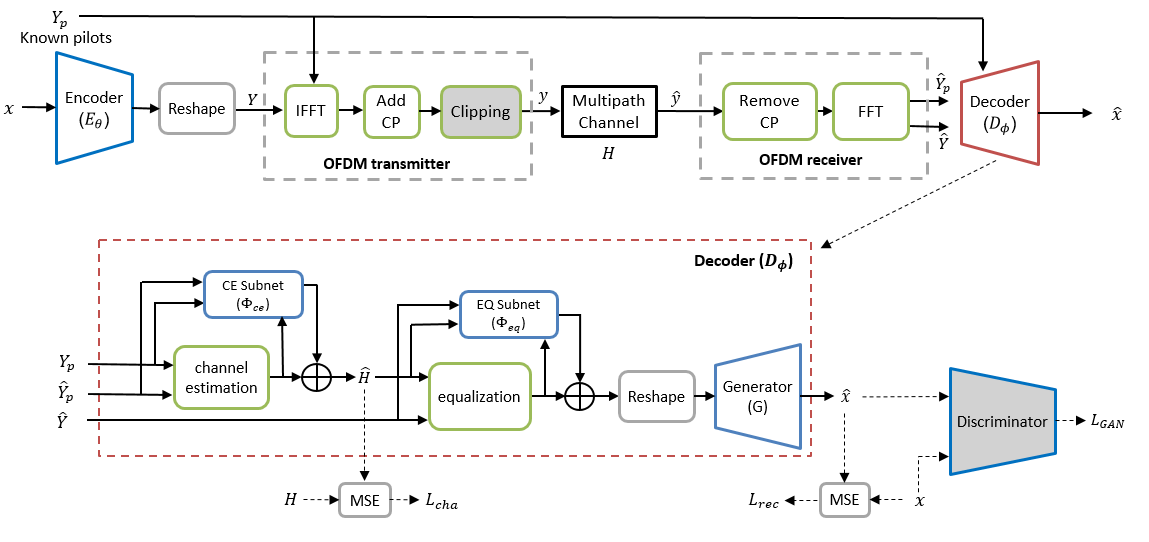}
    \caption{Top: Flow diagram for the proposed Deep JSCC framework with OFDM extension. Bottom: The detailed decoder structure proposed in this paper. We introduce explicit channel estimation and equalization methods as domain knowledge and enhance them with two additional residual connections. Blue boxes represent the networks to be trained. Green boxes denote domain knowledge operations that are fixed during training. The Clipping and Discriminator with gray background are optional extensions to our model.}
    \label{fig:fig_ofdm}
  \end{figure*}
}

\newcommand{\fignet}{
  \begin{figure}[t]
    \centering
    \includegraphics[width=\columnwidth]{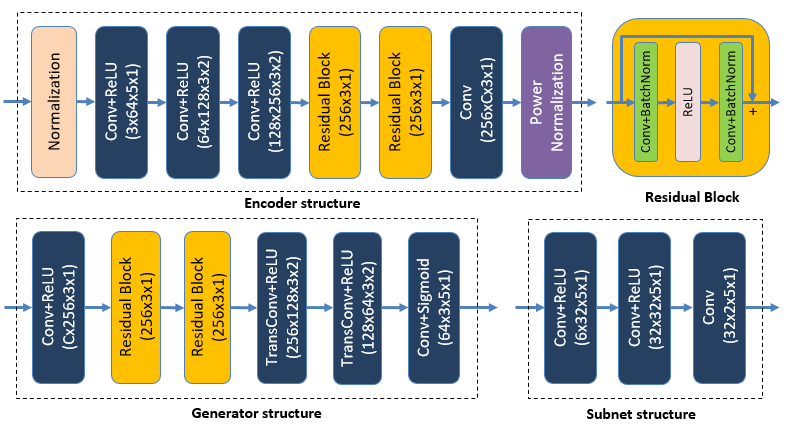}
    \caption{Network structure for the proposed method. There is a batch normalization layer between each convolutional layer and activation function (although not shown in the figure). The parameters for blue boxes are in the format of \textit{input channel size $\times$ output channel size $\times$ kernel size $\times$ stride}. The parameters for residual blocks are in the format of \textit{channel size $\times$  kernel size $\times$ stride}. }
    \label{fig:fig_net}
  \end{figure}
}

\newcommand{\figfeedback}{
  \begin{figure}[t]
    \centering
    \includegraphics[width=\columnwidth]{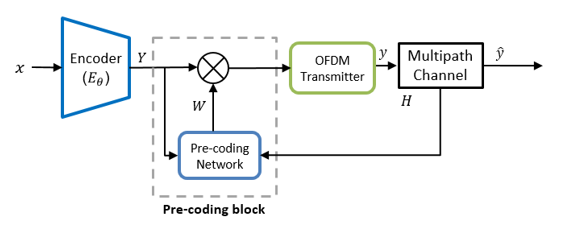}
    \caption{Encoder structure with CSI feedback.}
    \label{fig:fig_feedback}
  \end{figure}
}

\newcommand{\figresult}{
  \begin{figure*}[t]
    \centering
    \begin{subfigure}{0.49\linewidth}
       \centering\includegraphics[width=230pt]{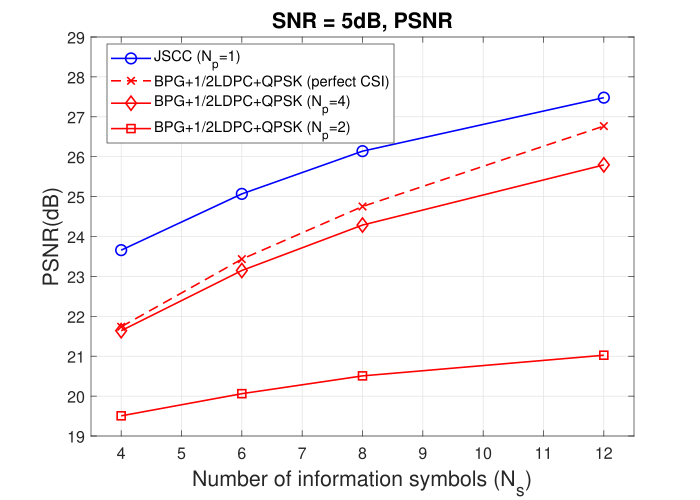}
       \caption{\label{fig:fig_psnr1}}
    \end{subfigure}%
    \begin{subfigure}{0.49\linewidth}
       \centering\includegraphics[width=230pt]{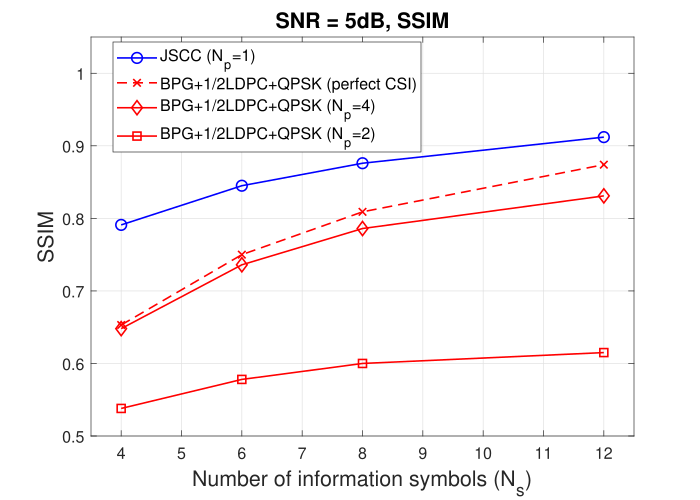}
       \caption{\label{fig:fig_ssim1}}
    \end{subfigure}
    \begin{subfigure}{0.49\linewidth}
       \centering\includegraphics[width=230pt]{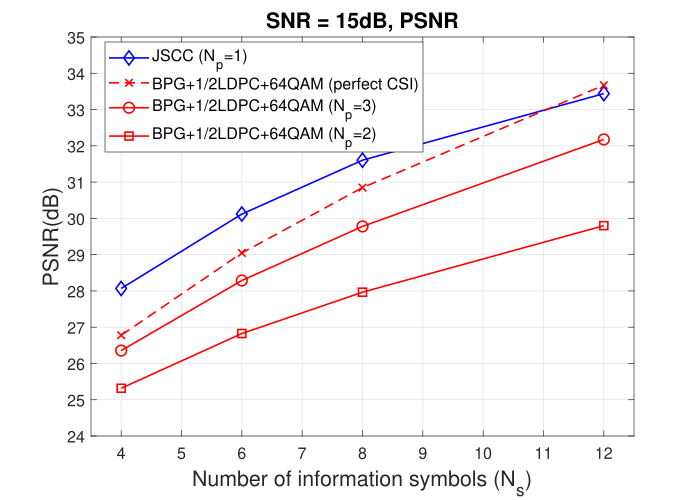}
       \caption{\label{fig:fig_psnr2}}
    \end{subfigure}
    \begin{subfigure}{0.49\linewidth}
       \centering\includegraphics[width=230pt]{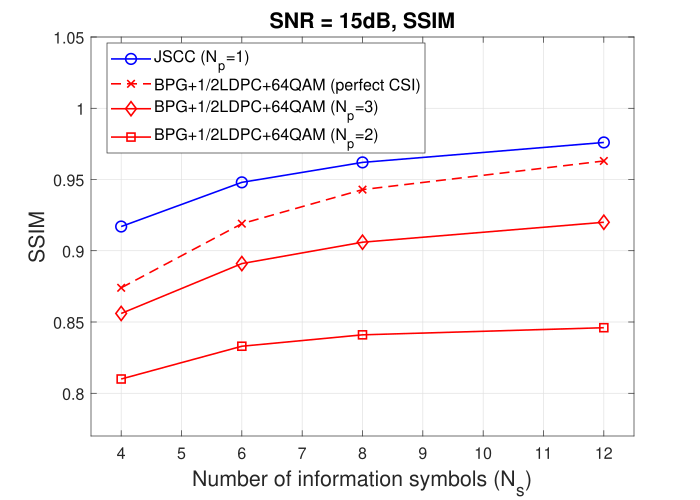}
       \caption{\label{fig:fig_ssim2}}
    \end{subfigure}
    
\caption{Performance of the proposed method with respect to the number of information symbols ($N_s$) for a fixed SNR of 5dB (a)(b) and 15dB (c)(d). Image quality is evaluated by PSNR (a)(c) and SSIM (b)(d).}
\label{fig:fig_result1}
\end{figure*}
}

\newcommand{\figresultlarge}{
  \begin{figure*}[t]
    
    \centering
    \begin{subfigure}[b]{0.49\linewidth}
       \centering\includegraphics[width=230pt]{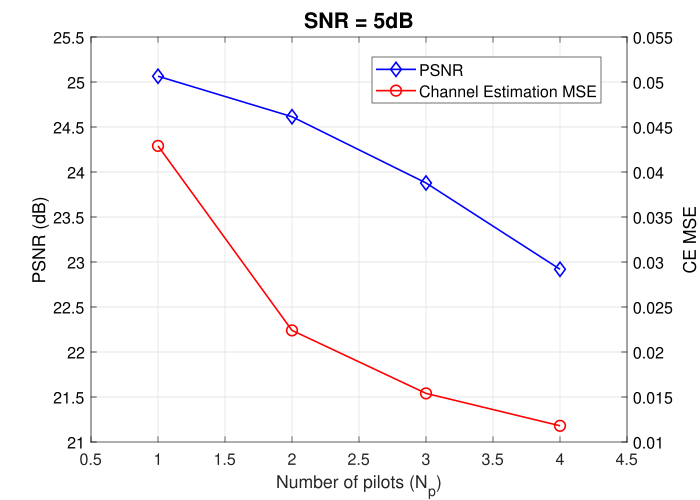}
       \caption{\label{fig:fig_mse1}}
    \end{subfigure}
    \begin{subfigure}[b]{0.49\linewidth}
       \centering\includegraphics[width=230pt]{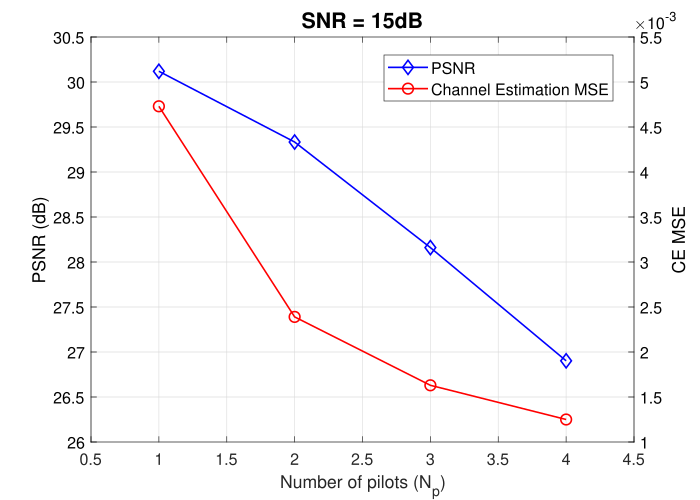}
       \caption{\label{fig:fig_mse2}}
    \end{subfigure}
\caption{Performance of the proposed method with respect to the number of pilot symbols ($N_p$) given a fixed transmission budget ($N_s+N_p=7$, $\text{CPP}=0.182$) at 5dB (a) and 15dB (b) SNR.}
\label{fig:fig_result2}
\end{figure*}
}

\newcommand{\figresultsnr}{
  \begin{figure*}[t]
    \centering
    \begin{subfigure}[b]{0.49\linewidth}
       \centering\includegraphics[width=230pt]{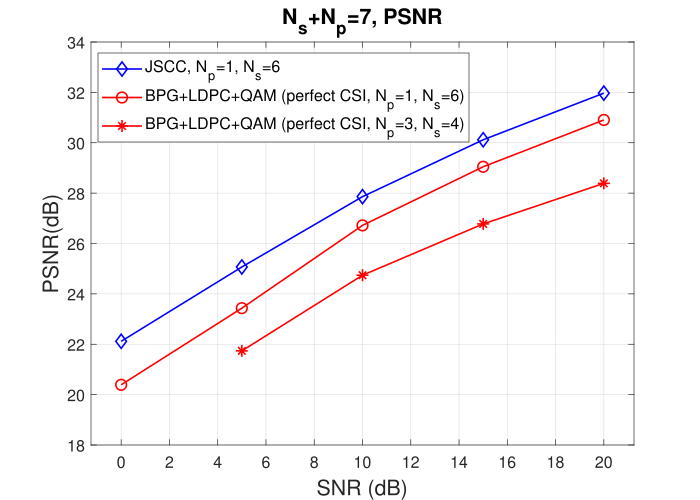}
       \caption{\label{fig:figsnr}}
    \end{subfigure}
    \begin{subfigure}[b]{0.49\linewidth}
       \centering\includegraphics[width=230pt]{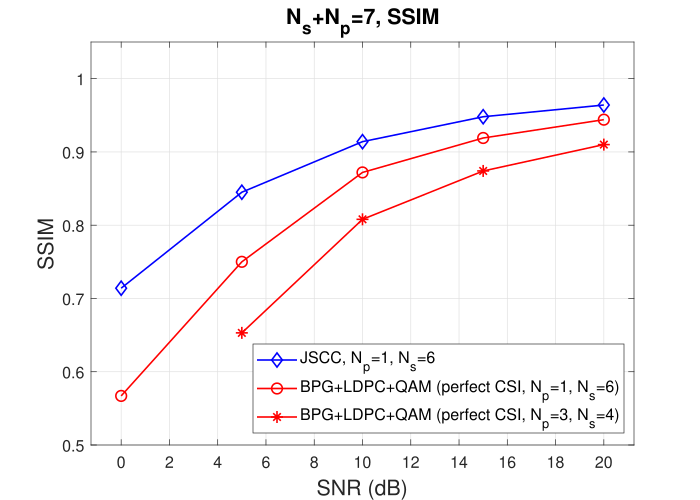}
       \caption{\label{fig:figssim}}
    \end{subfigure}
    \vspace{-0.3cm}
    \caption{Performance of the proposed method on CIFAR-10 with respect to SNR with fixed transmission budget ($N_s+N_p=7$, $\text{CPP}=0.182$). The baseline method uses perfect CSI regardless of $N_p$, while JSCC estimates CSI from a single pilot ($N_p=1$).}
    \label{fig:fig_result3}
  \end{figure*}
}

\newcommand{\figresultpapr}{
  \begin{figure}[t]
    \centering
    \includegraphics[width=250pt]{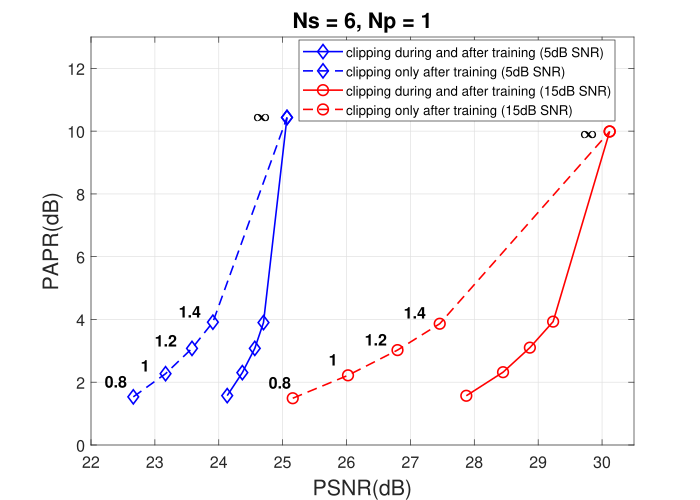}
    \caption{PAPR vs PSNR for different clipping ratios for both 5dB (blue) and 15dB (red) SNRs on CIFAR-10 dataset with $N_s=6, N_p=1$ ($\text{CPP}=0.182$). The corresponding clipping ratio is shown on the left of each marker. Solid line: network training is done with clipping. Dash line: clipping is applied after training while neural networks were not exposed to clipping during training.}
    \label{fig:fig_papr}
  \end{figure}
}

\newcommand{\figresultclip}{
  \begin{figure}[t]
    \centering
    \begin{subfigure}[b]{\columnwidth}
       \centering\includegraphics[width=230pt]{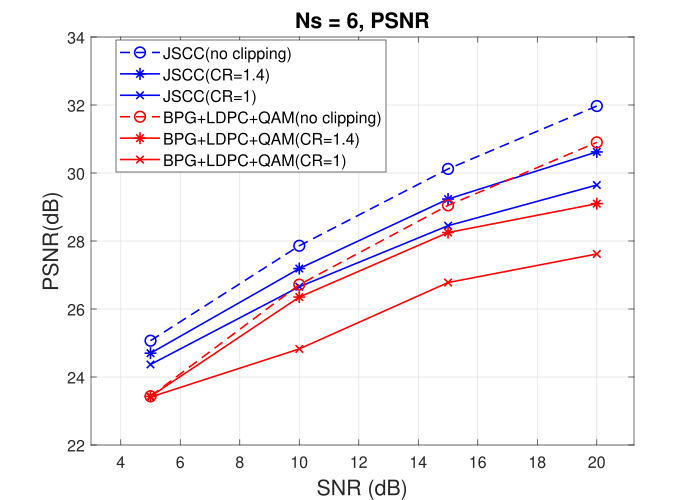}
       \caption{\label{fig:figsnr_clip}}
    \end{subfigure}
    \begin{subfigure}[b]{\columnwidth}
       \centering\includegraphics[width=230pt]{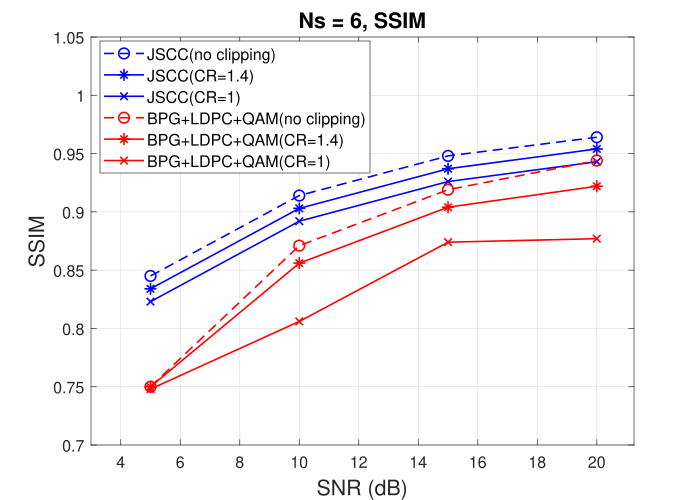}
       \caption{\label{fig:figssim_clip}}
    \end{subfigure}
    \vspace{-0.3cm}
    \caption{Performance of the proposed method on CIFAR-10 with respect to SNR for different clipping ratios with $N_s = 6, N_p = 1$ ($\text{CPP}=0.182$). The baseline method uses perfect CSI regardless of $N_p$.}
    \label{fig:fig_clip}
  \end{figure}
}

\newcommand{\figresultkodak}{
  \begin{figure*}[t]
    \captionsetup[subfigure]{labelformat=empty}
    \captionsetup{font={normalsize}}
    \begin{subfigure}[b]{0.25\linewidth}
       \centering
       \caption{\textbf{Original}}
       \vspace{-0.1cm}
       \includegraphics[width=115pt]{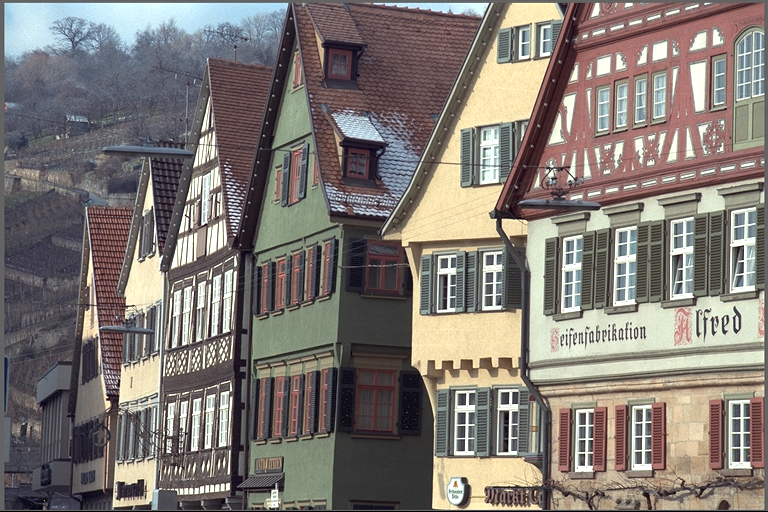}
       \vspace{-0.1cm}
       \caption{PSNR/SSIM/MS-SSIM}
    \end{subfigure}%
    \begin{subfigure}[b]{0.246\linewidth}
       \centering
       \caption{\textbf{BPG+LDPC}}
       \vspace{-0.1cm}
       \includegraphics[width=115pt]{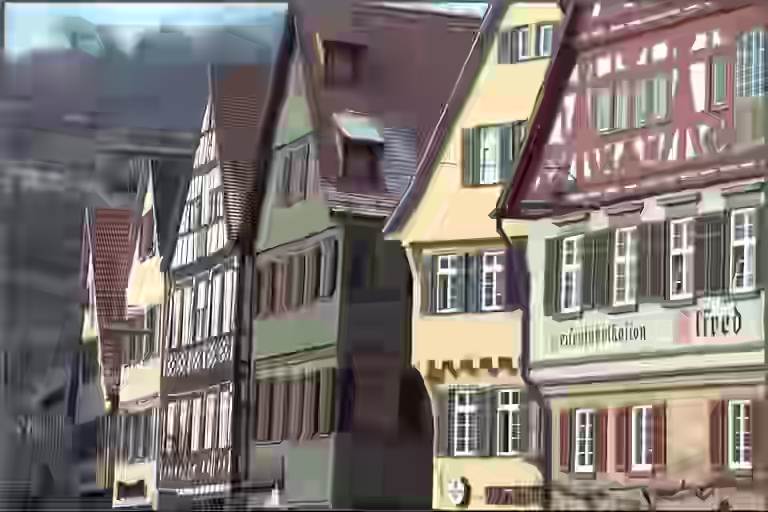}
       \vspace{-0.1cm}
       \caption{21.515/0.615/0.874}
    \end{subfigure}
    \begin{subfigure}[b]{0.246\linewidth}
       \centering
       \caption{\textbf{JSCC}}
       \vspace{-0.1cm}
       \includegraphics[width=115pt]{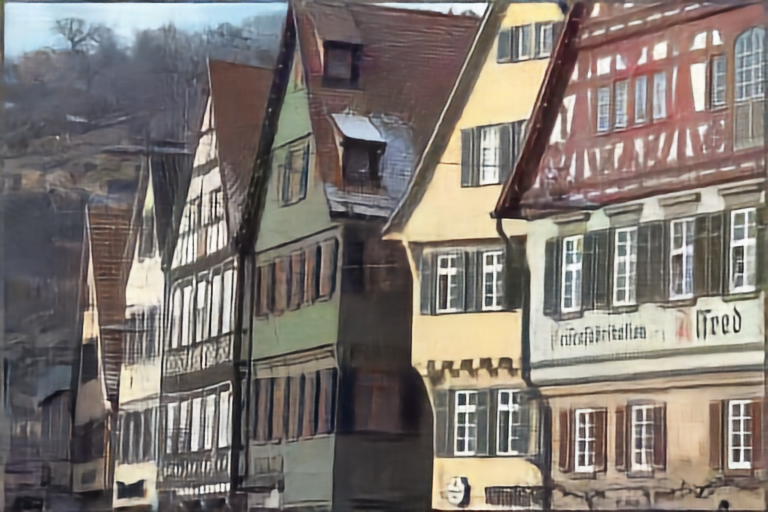}
       \vspace{-0.1cm}
       \caption{21.659/0.659/0.902}
    \end{subfigure}
    \begin{subfigure}[b]{0.246\linewidth}
       \centering
       \caption{\textbf{JSCC-\textit{adv}}}
       \vspace{-0.1cm}
       \includegraphics[width=115pt]{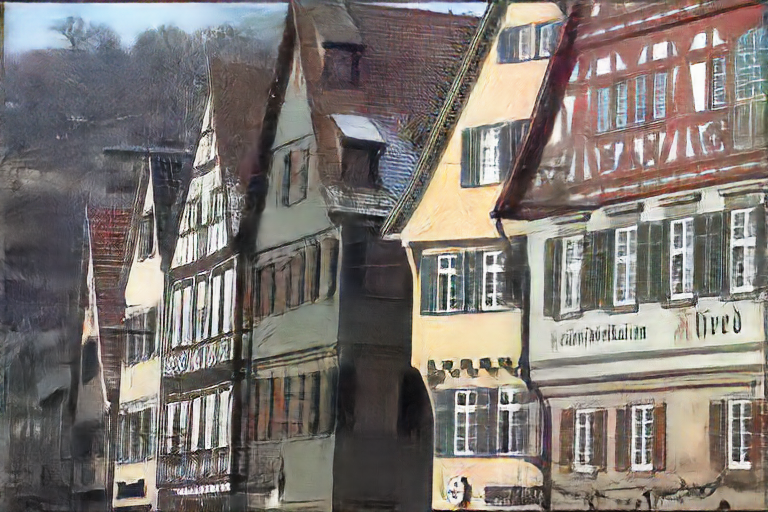}
       \vspace{-0.1cm}
       \caption{20.072/0.589/0.864}
    \end{subfigure}
    
    \captionsetup[subfigure]{labelformat=empty}
    \captionsetup{font={normalsize}}
    \begin{subfigure}[b]{0.25\linewidth}
       \centering
       \includegraphics[width=115pt]{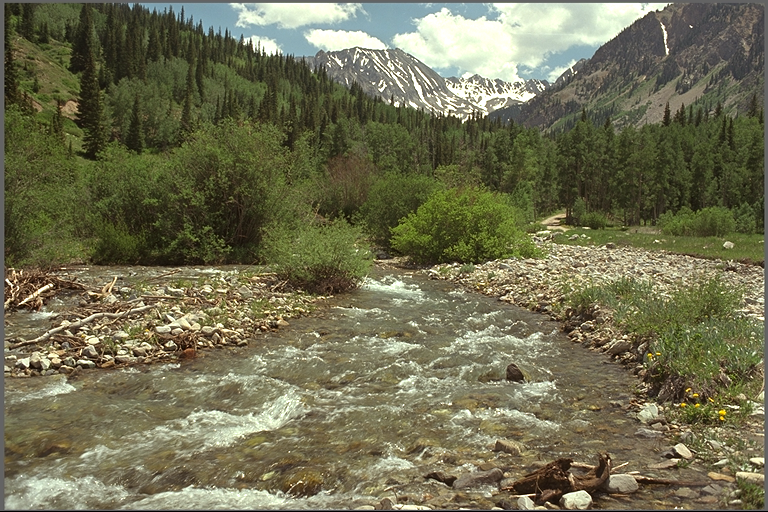}
       \vspace{-0.1cm}
       \caption{PSNR/SSIM/MS-SSIM}
    \end{subfigure}%
    \begin{subfigure}[b]{0.246\linewidth}
       \centering
       \includegraphics[width=115pt]{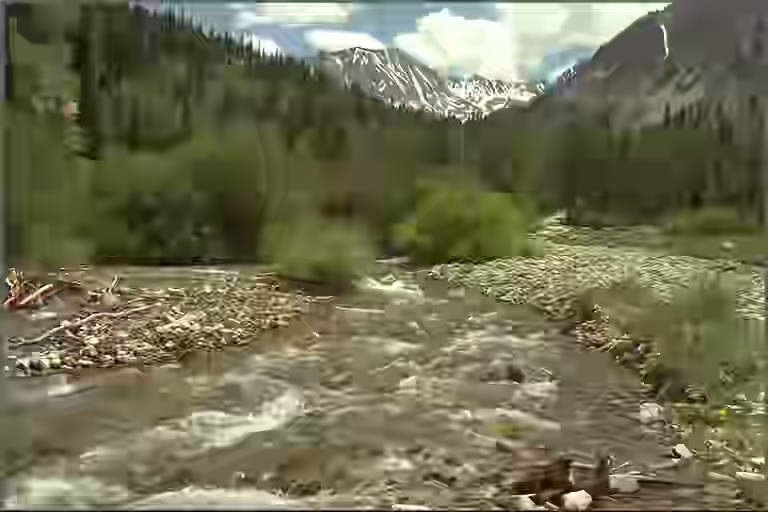}
       \vspace{-0.1cm}
       \caption{21.092/0.42/0.772}
    \end{subfigure}
    \begin{subfigure}[b]{0.246\linewidth}
       \centering
       \includegraphics[width=115pt]{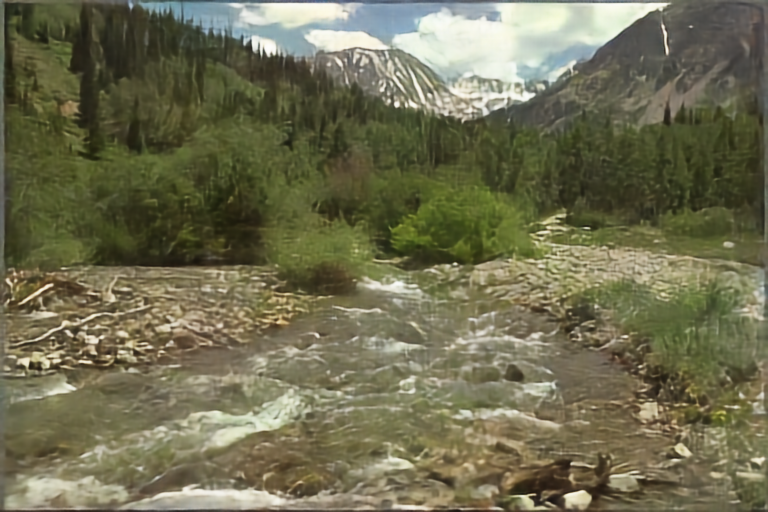}
       \vspace{-0.1cm}
       \caption{21.175/0.505/0.844}
    \end{subfigure}
    \begin{subfigure}[b]{0.246\linewidth}
       \centering
       \includegraphics[width=115pt]{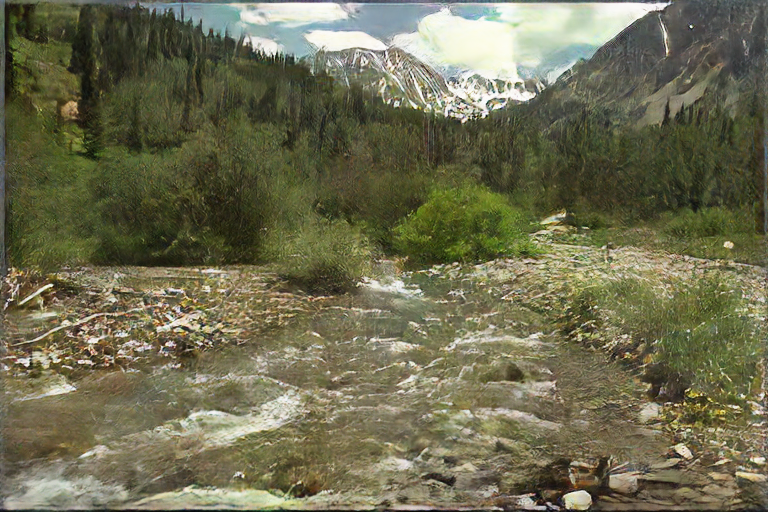}
       \vspace{-0.1cm}
       \caption{20.208/0.464/0.821}
    \end{subfigure}
    
    \captionsetup[subfigure]{labelformat=empty}
    \captionsetup{font={normalsize}}
    \begin{subfigure}[b]{0.25\linewidth}
       \centering
       \includegraphics[width=115pt]{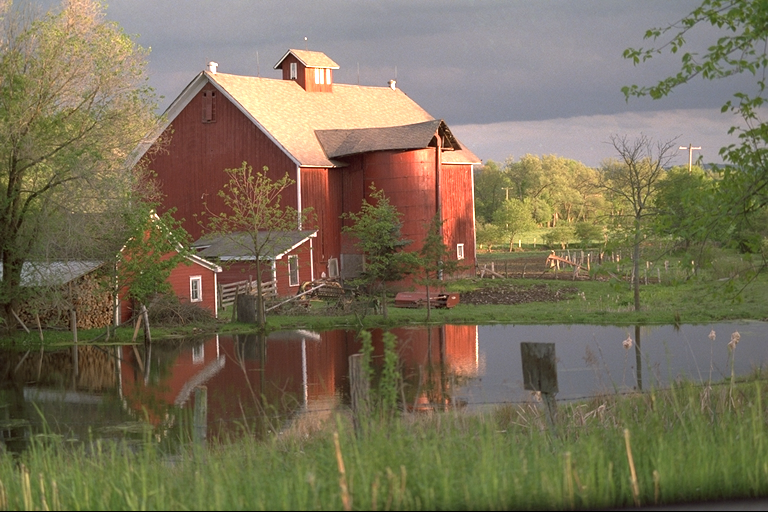}
       \vspace{-0.1cm}
       \caption{PSNR/SSIM/MS-SSIM}
    \end{subfigure}%
    \begin{subfigure}[b]{0.246\linewidth}
       \centering
       \includegraphics[width=115pt]{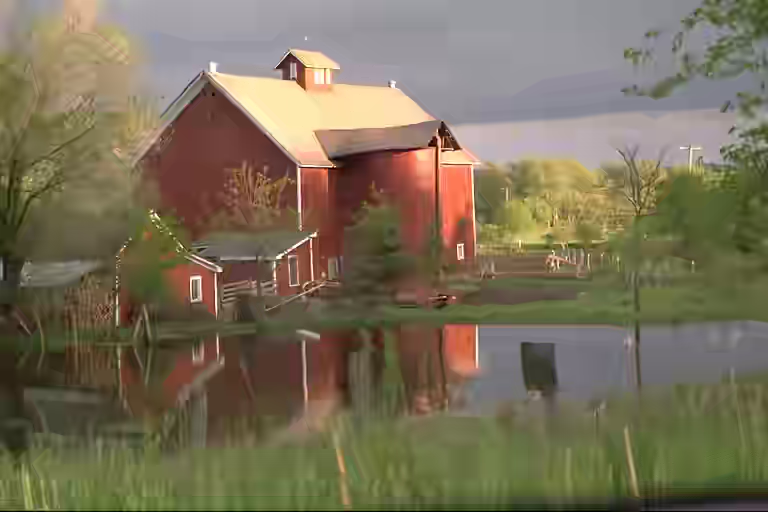}
       \vspace{-0.1cm}
       \caption{27.334/0.67/0.867}
    \end{subfigure}
    \begin{subfigure}[b]{0.246\linewidth}
       \centering
       \includegraphics[width=115pt]{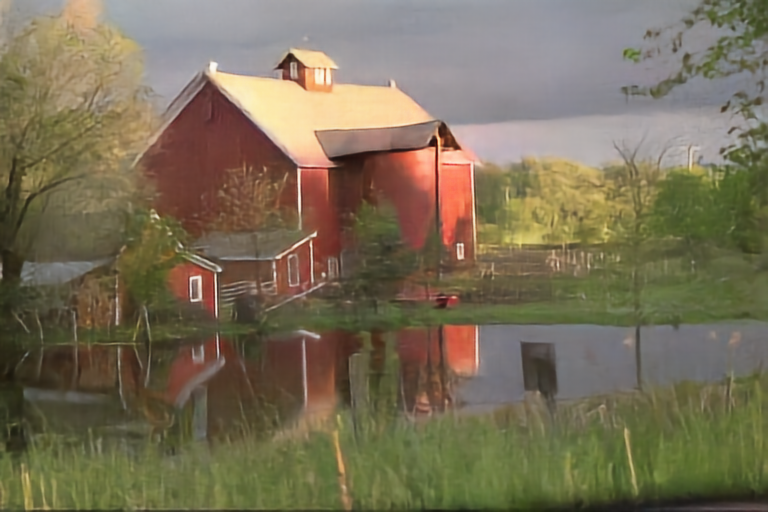}
       \vspace{-0.1cm}
       \caption{27.057/0.709/0.895}
    \end{subfigure}
    \begin{subfigure}[b]{0.246\linewidth}
       \centering
       \includegraphics[width=115pt]{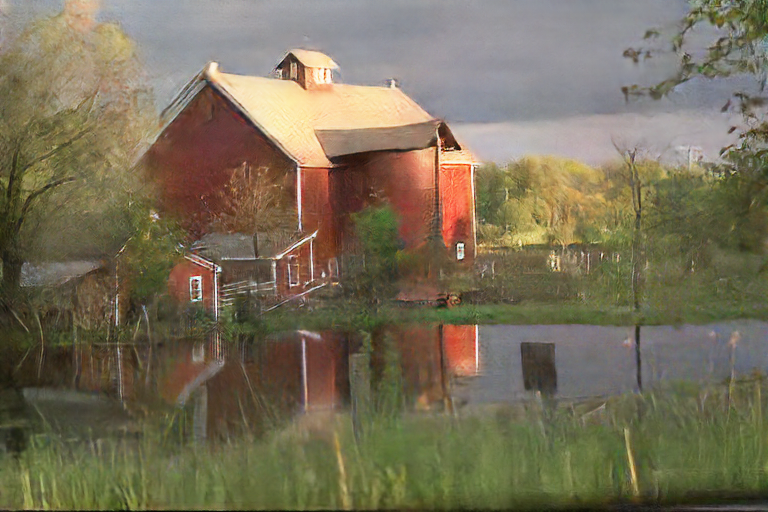}
       \vspace{-0.1cm}
       \caption{25.75/0.655/0.86}
    \end{subfigure}
    
    \captionsetup[subfigure]{labelformat=empty}
    \captionsetup{font={normalsize}}
    \begin{subfigure}[b]{0.25\linewidth}
       \centering
       \includegraphics[width=115pt]{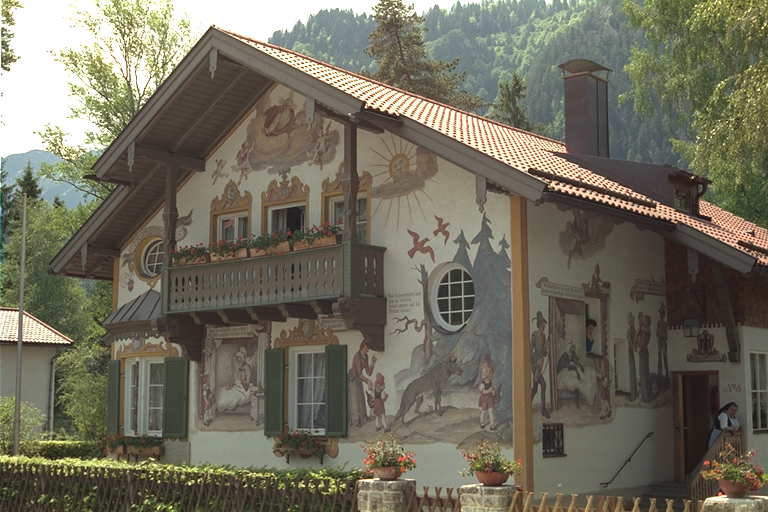}
       \vspace{-0.1cm}
       \caption{PSNR/SSIM/MS-SSIM}
    \end{subfigure}%
    \begin{subfigure}[b]{0.246\linewidth}
       \centering
       \includegraphics[width=115pt]{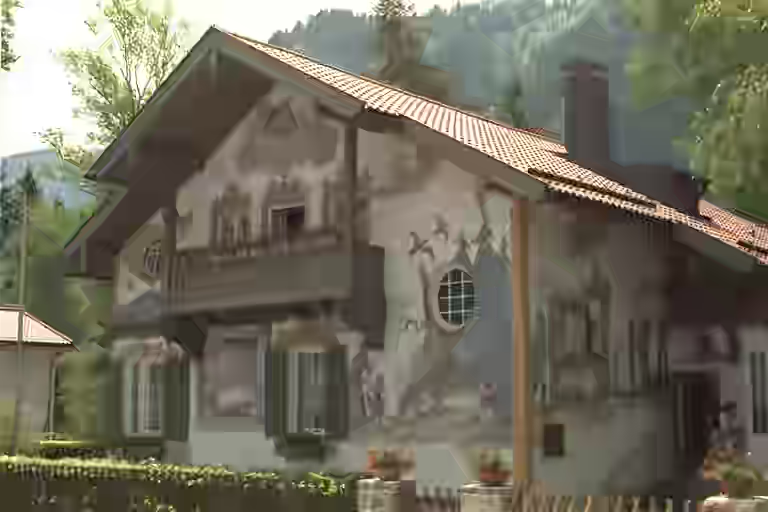}
       \vspace{-0.1cm}
       \caption{24.089/0.632/0.857}
    \end{subfigure}
    \begin{subfigure}[b]{0.246\linewidth}
       \centering
       \includegraphics[width=115pt]{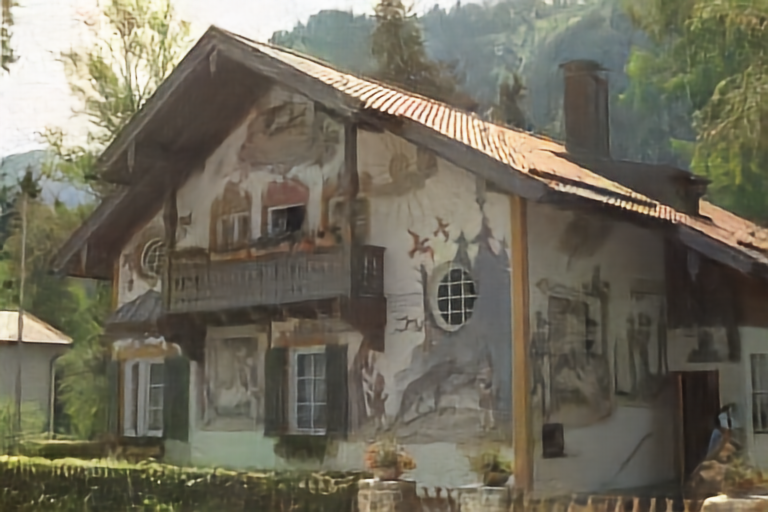}
       \vspace{-0.1cm}
       \caption{23.68/0.687/0.9}
    \end{subfigure}
    \begin{subfigure}[b]{0.246\linewidth}
       \centering
       \includegraphics[width=115pt]{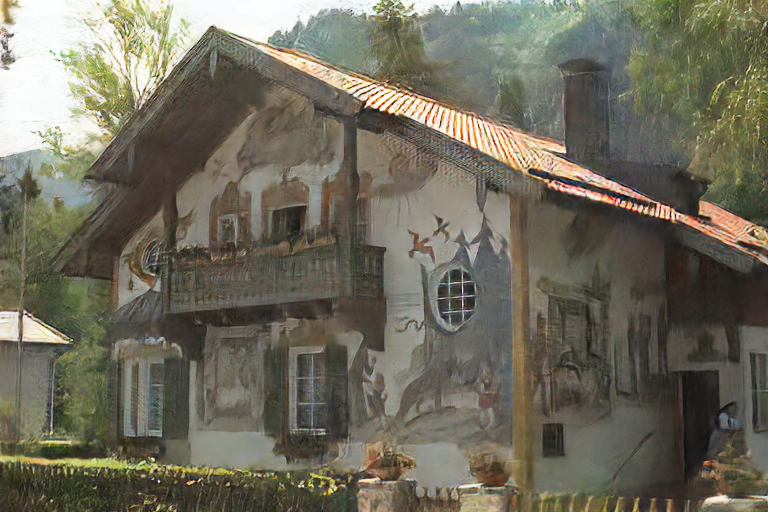}
       \vspace{-0.1cm}
       \caption{22.607/0.631/0.874}
    \end{subfigure}
    
    \captionsetup[subfigure]{labelformat=empty}
    \captionsetup{font={normalsize}}
    \begin{subfigure}[b]{0.25\linewidth}
       \centering
       \includegraphics[width=115pt]{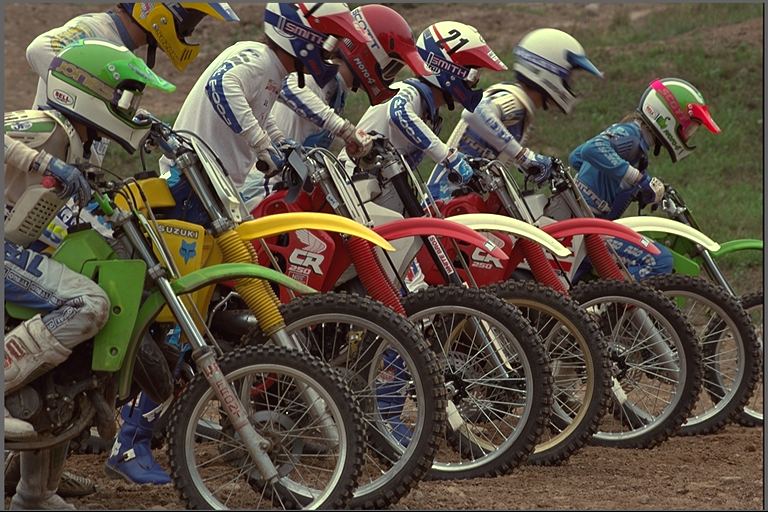}
       \vspace{-0.1cm}
       \caption{PSNR/SSIM/MS-SSIM}
    \end{subfigure}%
    \begin{subfigure}[b]{0.246\linewidth}
       \centering
       \includegraphics[width=115pt]{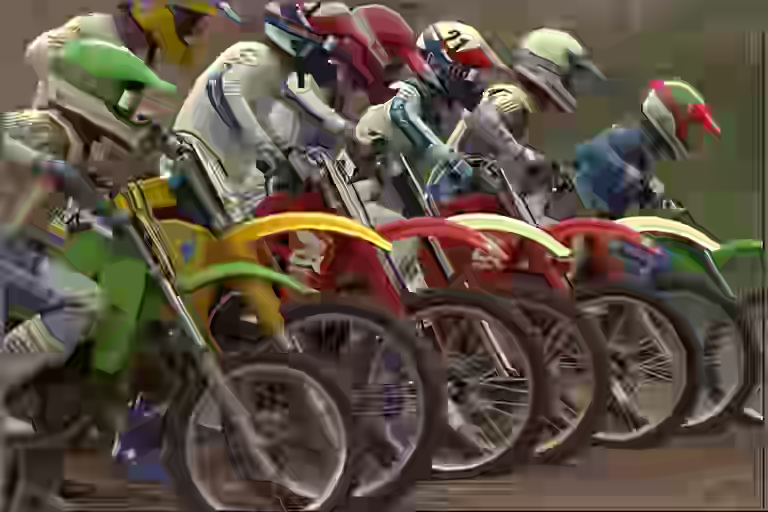}
       \vspace{-0.1cm}
       \caption{21.534/0.522/0.817}
    \end{subfigure}
    \begin{subfigure}[b]{0.246\linewidth}
       \centering
       \includegraphics[width=115pt]{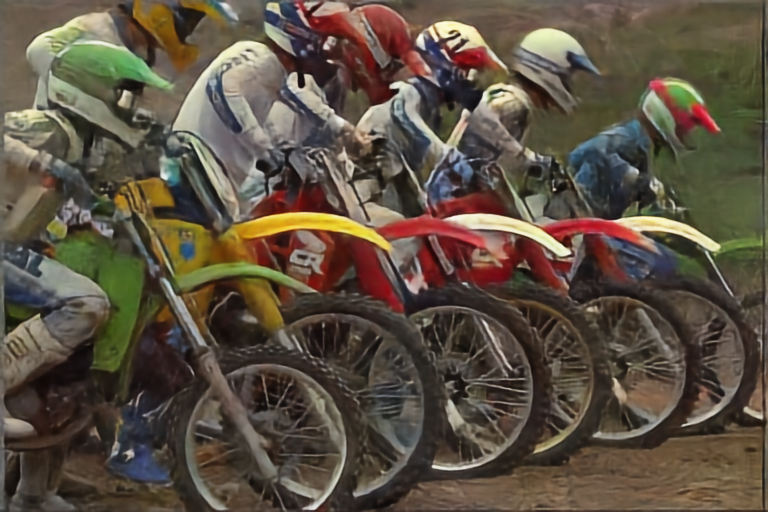}
       \vspace{-0.1cm}
       \caption{22.512/0.636/0.882}
    \end{subfigure}
    \begin{subfigure}[b]{0.246\linewidth}
       \centering
       \includegraphics[width=115pt]{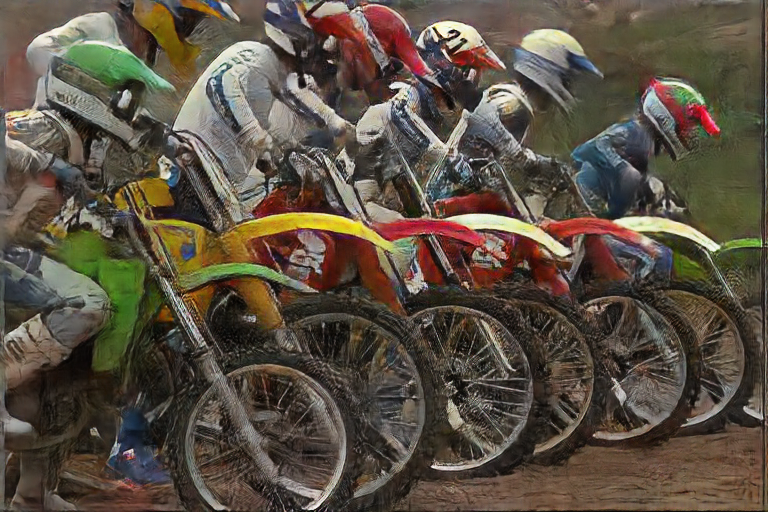}
       \vspace{-0.1cm}
       \caption{21.059/0.56/0.849}
    \end{subfigure}

\captionsetup{font={small}}   
\caption{Visual comparison between images achieved from the baseline separated coding schemes and the proposed JSCC method on Kodak, with $\text{CPP}=0.039$. The performance is associated with each recovered image in a format of PSNR(dB)/SSIM/MS-SSIM. }
\label{fig:fig_result_kodak}
\end{figure*}
}

\newcommand{\figrob}{
 \begin{figure}[t]
    \centering
    \begin{subfigure}[b]{\columnwidth}
       \centering\includegraphics[width=230pt]{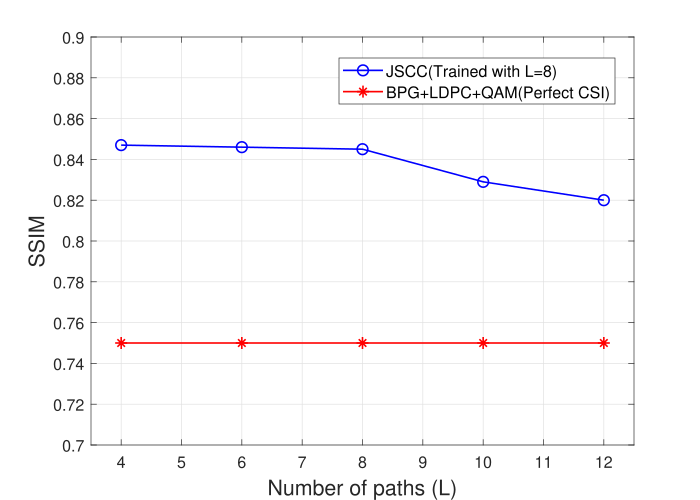}
       \caption{\label{fig:figssim_rob1}}
    \end{subfigure}
    \begin{subfigure}[b]{\columnwidth}
       \centering\includegraphics[width=230pt]{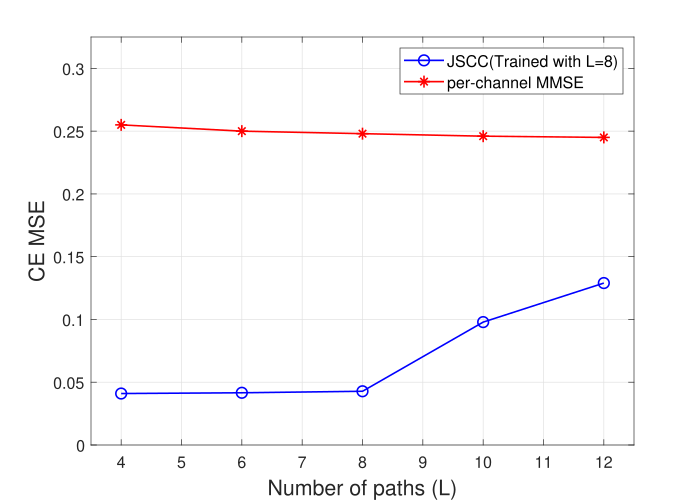}
       \caption{\label{fig:figssim_rob2}}
    \end{subfigure}
    \caption{Robustness test for multipath channel model parameter ($L$) mismatch scenarios. The JSCC is trained with $N_s=6$, $N_p=1$ ($\text{CPP}=0.182$), and 5dB SNR. (a): SSIM, (b): channel estimation MSE. }
    \label{fig:fig_rob}
  \end{figure}
}

\newcommand{\figrobsnr}{
 \begin{figure}[t]
    \centering
    
    \begin{subfigure}[b]{\columnwidth}
       \centering\includegraphics[width=230pt]{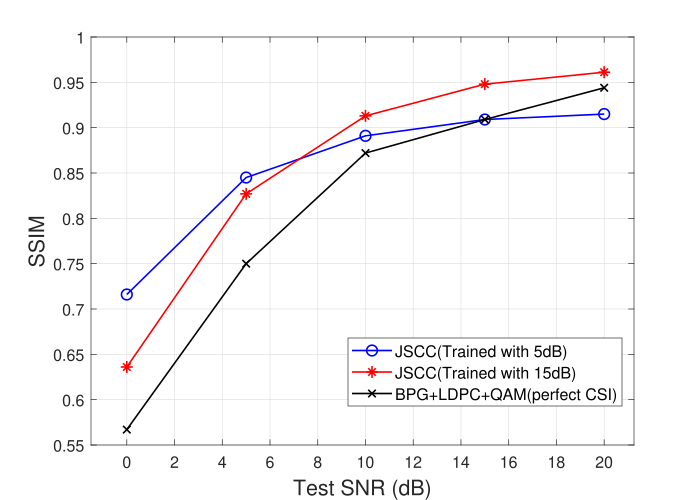}
       \caption{\label{fig:figssim_rob3}}
    \end{subfigure}
    \begin{subfigure}[b]{\columnwidth}
       \centering\includegraphics[width=230pt]{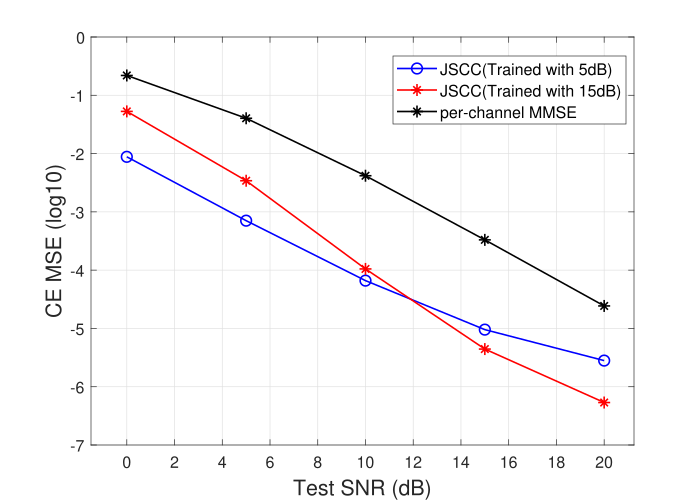}
       \caption{\label{fig:figssim_rob4}}
    \end{subfigure}
    \caption{Robustness test for training vs. testing SNR mismatch scenarios. The JSCC is trained with CIFAR-10 images for $N_s=6$ and $N_p=1$ ($\text{CPP}=0.182$). (a): SSIM, (b): channel estimation MSE}
    \label{fig:fig_robsnr}
  \end{figure}
}

\newcommand{\figlarge}{
 \begin{figure}[t]
    \centering
    \begin{subfigure}[b]{\columnwidth}
       \centering\includegraphics[width=230pt]{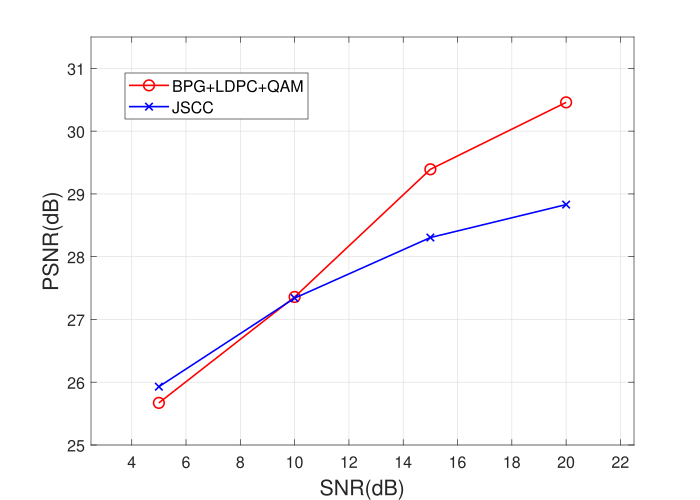}
       \caption{\label{fig:figssim_large1}}
    \end{subfigure}
    \begin{subfigure}[b]{\columnwidth}
       \centering\includegraphics[width=230pt]{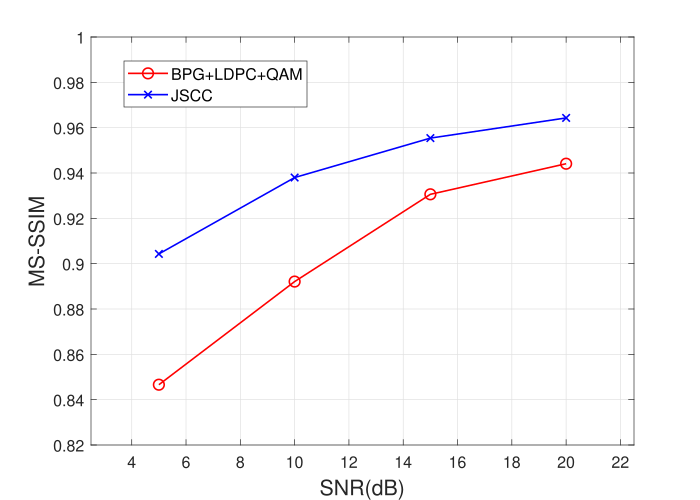}
       \caption{\label{fig:figssim_large2}}
    \end{subfigure}
    \caption{Performance of the proposed method on Kodak dataset. (a): PSNR, (b): MS-SSIM. }
    \label{fig:fig_kodak_result}
  \end{figure}
}

\newcommand{\figfeedbackresult}{
 \begin{figure*}[t]
    \centering
    \begin{subfigure}[b]{\columnwidth}
       \centering\includegraphics[width=230pt]{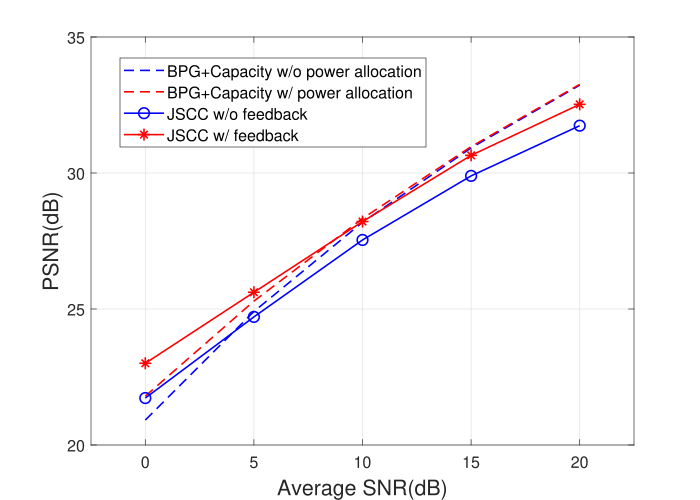}
       \caption{\label{fig:figssim_large3}}
    \end{subfigure}
    \begin{subfigure}[b]{\columnwidth}
       \centering\includegraphics[width=230pt]{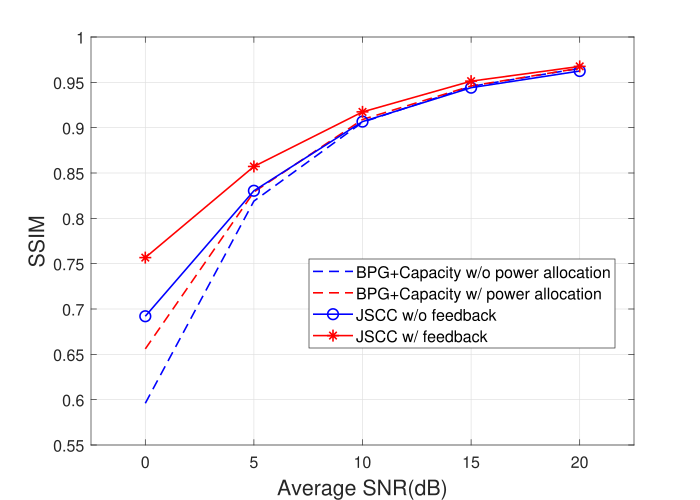}
       \caption{\label{fig:figssim_large4}}
    \end{subfigure}
    \caption{Performance of the proposed method with perfect CSI feedback on CIFAR-10 with respect to SNR.  }
    \label{fig:fig_feedback_result}
  \end{figure*}
}

\newcommand{\fignetresult}{
 \begin{figure}[th]
    \centering
    \begin{subfigure}[b]{\columnwidth}
       \centering\includegraphics[width=220pt]{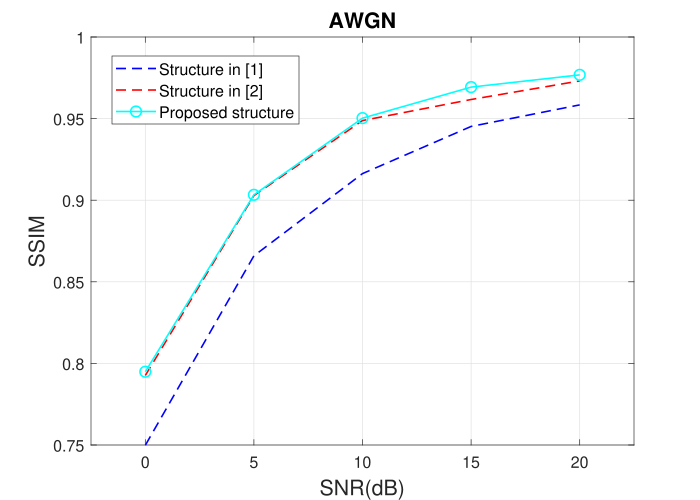}
       \caption{\label{fig:figssim_net1}}
    \end{subfigure}
    \begin{subfigure}[b]{\columnwidth}
       \centering\includegraphics[width=220pt]{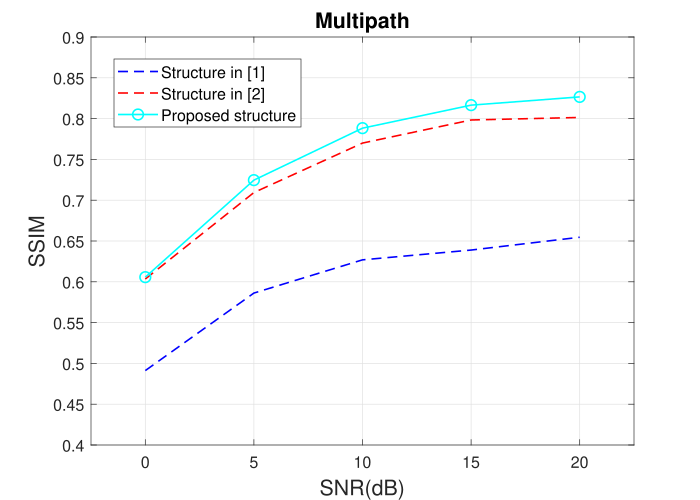}
       \caption{\label{fig:figssim_net2}}
    \end{subfigure}
    \caption{Comparison of different neural network structures in (a) AWGN and (b) Multipath channel  }
    \label{fig:fig_net_result}
  \end{figure}
}

\newcommand{\tablelow}{
\begin{table*}[tbp]
\small
\caption{Evaluation of different components (SNR=5dB)}
\begin{center}
\begin{tabular}{c|ccc|ccc|ccc}
\hline
\textbf{Dataset}&\multicolumn{3}{c|}{\textbf{CIFAR-10, CPP=0.21}} &\multicolumn{3}{c|}{\textbf{CIFAR-100, CPP=0.21}} &\multicolumn{3}{c}{\textbf{CelebA, CPP=0.05}} \\
\cline{1-1} 
\textbf{Metric} & \textbf{PSNR}& \textbf{SSIM}& \textbf{CE MSE} & \textbf{PSNR}& \textbf{SSIM}& \textbf{CE MSE} & \textbf{PSNR}& \textbf{SSIM}& \textbf{CE MSE} \\
\hline\hline
Direct w/o OFDM &23.493&0.791&N/A&23.324&0.766&N/A&23.453&0.743&N/A\\
OFDM w/o CE\&EQ &24.611&0.836&N/A&24.534&0.821&N/A&24.733&0.798&N/A\\
OFDM+CE+EQ &24.438&0.824&0.25&25.36& 0.809&0.249& 24.661 & 0.793& 0.247\\
OFDM+CE+EQ+$\Phi_{ce}$ &25.539&0.86&0.0429&25.463&0.847&0.0427&25.552 & 0.821& 0.0426\\
OFDM+CE+EQ+$\Phi_{ce}$+$\Phi_{eq}$ & \textbf{25.626}& \textbf{0.862}& 0.0429& \textbf{25.567} & \textbf{0.85}&0.0428 &\textbf{25.673} & \textbf{0.825}& 0.0427 \\
\hline
\end{tabular}
\label{tab1}
\end{center}
\end{table*}
}

\newcommand{\tablehigh}{
\begin{table*}[tbp]
\small
\caption{Evaluation of different components (SNR=15dB)}
\begin{center}
\begin{tabular}{c|ccc|ccc|ccc}
\hline
\textbf{Dataset}&\multicolumn{3}{c|}{\textbf{CIFAR-10, CPP=0.21}} &\multicolumn{3}{c|}{\textbf{CIFAR-100, CPP=0.21}} &\multicolumn{3}{c}{\textbf{CelebA, CPP=0.05}} \\
\cline{1-1} 
\textbf{Metric} & \textbf{PSNR}& \textbf{SSIM}& \textbf{CE MSE} & \textbf{PSNR}& \textbf{SSIM}& \textbf{CE MSE} & \textbf{PSNR}& \textbf{SSIM}& \textbf{CE MSE} \\
\hline\hline
Direct w/o OFDM &25.604&0.857&N/A&25.47&0.839&N/A&25.298&0.8&N/A\\
OFDM w/o CE\&EQ &29.361&0.941&N/A&29.241&0.933&N/A&28.462&0.896&N/A\\
OFDM+CE+EQ &29.848&0.945&0.0309&29.731&0.937&0.0309& 28.817 & 0.901& 0.0307\\
OFDM+CE+EQ+$\Phi_{ce}$ &30.796&0.955&0.0047&30.657&0.948&0.0047&29.463 & 0.912& 0.0047\\
OFDM+CE+EQ+$\Phi_{ce}$+$\Phi_{eq}$ & \textbf{30.905}& \textbf{0.956}& 0.0047& \textbf{30.686} & \textbf{0.949}&0.0047 &\textbf{29.587} & \textbf{0.915}& 0.0047 \\
\hline
\end{tabular}
\label{tab2}
\end{center}
\end{table*}
}

\newcommand{\tablecomplex}{
\begin{table}[tbp]
\small
\caption{Complexity Analysis for CIFAR-10}
\begin{center}
\begin{tabular}{c|c|c}
\textbf{Metric} & Parameters (M) & FLOPS (M)\\
\hline\hline
Direct w/o OFDM & 5.5631 & 242.5\\
OFDM w/o CE\&EQ & 5.5469 &241.47\\
OFDM+CE+EQ & 5.5354 & 240.73\\
OFDM+CE+EQ+$\Phi_{ce}$ & 5.5675 & 242.78\\
OFDM+CE+EQ+$\Phi_{ce}$+$\Phi_{eq}$ & 5.5997 & 244.83\\
\hline
\end{tabular}
\label{tab3}
\end{center}
\end{table}
}

\newcommand{\tableadv}{

\begin{table}[tbp]
\footnotesize
\caption{Evaluation of Adversarial Loss on CIFAR-10}
\begin{center}
\begin{tabular}{c|ccc|ccc}
\hline
\textbf{SNR}&\multicolumn{3}{c|}{\textbf{5dB}} &\multicolumn{3}{c}{\textbf{15dB}} \\
\cline{1-1} 
\textbf{Metric} & \textbf{PSNR}& \textbf{SSIM}& \textbf{CA} & \textbf{PSNR}& \textbf{SSIM}& \textbf{CA} \\
\hline\hline
Original & $\infty$ & 1 & 93.62\% & $\infty$ & 1 & 93.62\% \\
BPG+LDPC &23.433&0.75&32.3\%&29.047&0.919&81.9\%\\
JSCC &\textbf{25.068}&\textbf{0.845}&64.42\%&\textbf{30.126}&\textbf{0.948}&87.21\%\\
JSCC-\textit{adv} &24.129&0.826&\textbf{74.22\%}&29.338&0.937&\textbf{87.74\%}\\

\hline
\end{tabular}
\label{tab_adv}
\end{center}
\end{table}

}

\newcommand{\figvisual}{
 \begin{figure}[t]
    \centering
    
    \begin{subfigure}[b]{\columnwidth}
       \centering\includegraphics[width=250pt]{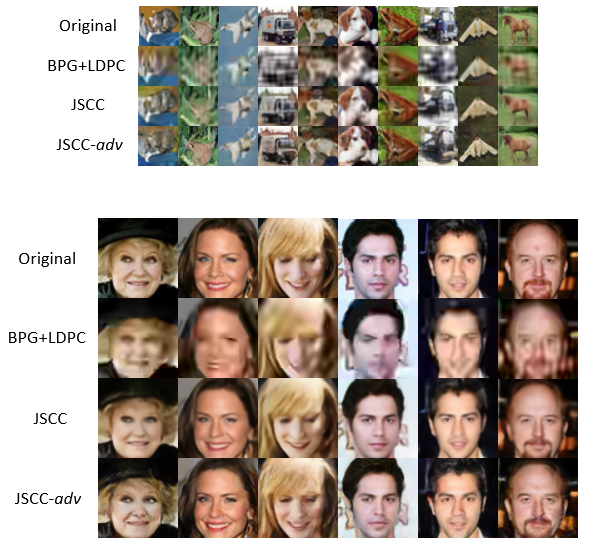}
    \end{subfigure}
    \caption{Visualizations of CIFAR-10 (top, $\text{CPP}=0.182$) and CelebA (bottom, $\text{CPP}=0.111$). }
    \label{fig:fig_visual}
  \end{figure}
}

\newcommand{\tableabc}{
\begin{table}[tbp]
\scriptsize
\caption{Ablation Studies on $\lambda_c$}
\begin{center}
\begin{tabular}{c|c|>{\centering\arraybackslash}p{0.6cm}|>{\centering\arraybackslash}p{0.6cm}|>{\centering\arraybackslash}p{0.6cm}|>{\centering\arraybackslash}p{0.6cm}|>{\centering\arraybackslash}p{0.6cm}|>{\centering\arraybackslash}p{0.6cm}}
 & w/o $\Phi_{ce}$ &\multicolumn{6}{c}{with $\Phi_{ce}$} \\
 \cline{2-8} 
$\lambda_c$ & N/A & 0 & 0.05 & 0.1 & 0.5 & 1 & 5\\
\hline\hline
PSNR (dB) & 23.865 &25.036 & 25.045 & 25.061 & 25.068 & 25.072 & 25.069\\
CE MSE & 2.4969 & 0.9554  & 0.4449 & 0.4336 & 0.4282 & 0.4277 & 0.4275\\
\hline
\end{tabular}
\label{tablambdac}
\end{center}
\end{table}
}

\newcommand{\tableabg}{
\begin{table}[tbp]
\scriptsize
\caption{Ablation Studies on $\lambda_g$}
\begin{center}
\begin{tabular}{c|c|c|c|c|c|c}
$\lambda_g$ & 0 & 2e-4 & 5e-4 & 2e-3 & 5e-3 & 0.02\\
\hline\hline
PSNR (dB) & 25.068 & 25.066 & 25.06 & 24.129 & 23.508 & 22.239\\
CA (\%) & 64.42 & 64.45 & 64.55 & 74.22 & 70.81 & 68.13\\
\hline
\end{tabular}
\label{tablambdag}
\end{center}
\end{table}
}

\newcommand{\figvisuallambda}{
 \begin{figure}[t]
    \centering
    
    \begin{subfigure}[b]{0.9\columnwidth}
       \centering\includegraphics[width=230pt]{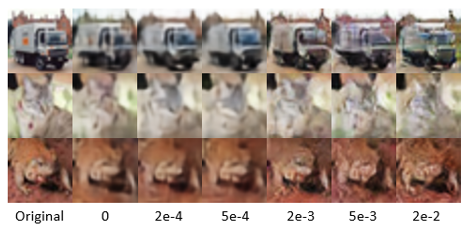}
    \end{subfigure}
    \caption{Visualizations of CIFAR-10 with different $\lambda_g$}
    \label{fig:fig_visual_lambdag}
  \end{figure}
}

\begin{document}
%
\title{OFDM-guided Deep Joint Source Channel Coding for Wireless Multipath Fading Channels}
%
%
%

\author{Mingyu~Yang,~\IEEEmembership{Student Member,~IEEE,}
        Chenghong~Bian,
        and~Hun-Seok~Kim,~\IEEEmembership{Member,~IEEE}
\thanks{M. Yang, C. Bian and H. Kim are with the Department
of Electrical and Computer Engineering, University of Michigan, Ann Arbor,
MI, 48109 USA e-mail: (mingyuy@umich.edu; chbian@umich.edu; hunseok@umich.edu }
\thanks{This work was funded in part by DARPA YFA \#D18AP00076 and NSF CAREER \#1942806. }
\thanks{The source code is available at https://github.com/mingyuyng/OFDM-guided-JSCC. }
}

%
%

\markboth{Journal of \LaTeX\ Class Files,~Vol.~14, No.~8, August~2015}%
{Shell \MakeLowercase{\textit{et al.}}: Bare Demo of IEEEtran.cls for IEEE Communications Society Journals}
%



\maketitle

\begin{abstract}
We investigate joint source channel coding (JSCC) for wireless image transmission over multipath fading channels. Inspired by recent works on deep learning based JSCC and model-based learning methods, we combine an autoencoder with orthogonal frequency division multiplexing (OFDM) to cope with multipath fading. The proposed encoder and decoder use convolutional neural networks (CNNs) and directly map the source images to complex-valued baseband samples for OFDM transmission. The multipath channel and OFDM are represented by non-trainable (deterministic) but differentiable layers so that the system can be trained end-to-end. Furthermore, our JSCC decoder further incorporates explicit channel estimation, equalization, and additional subnets to enhance the performance. The proposed method exhibits 2.5 -- 4 dB SNR gain for the equivalent image quality compared to conventional schemes that employ state-of-the-art but separate source and channel coding such as BPG and LDPC. The performance further improves when the system incorporates the channel state information (CSI) feedback. 
The proposed scheme is robust against OFDM signal clipping and  parameter mismatch for the channel model used in training and evaluation. 
\end{abstract}


\section{Introduction}
%
%
%
%
\IEEEPARstart{S}{hannon's} separation theorem\cite{shannon1948mathematical} states that, if the minimum achievable source coding rate of a given source is less than the channel capacity, then the source can be reliably transmitted through the channel. Thus, the source and channel coding can be designed separately without loss of optimality. This famous theorem is the basis for the tremendous success of practical communication systems because of the modularity it affords. For example, most modern systems for wireless image transmission first compress the image with a source coding algorithm (e.g., JPEG, WebP, BPG) and then encode the bit stream with a source-independent channel code (e.g., LDPC, Polar, Turbo, etc.) as shown in Figure \ref{fig:fig_intro} top. 
\figintroJSCC
However, to reach the optimality of the separation theorem, we require infinite codeword length which is not practical for applications that impose low latency real-time communication and/or low computation complexity constraints such as autonomous driving and the Internet of Things. Moreover, when the channel quality falls below a certain threshold, the channel coding may not provide satisfying error corrections and the source coding will tend to fail catastrophically. As a result, it is beneficial to jointly optimize source coding and channel coding for relatively short messages, which is referred to as joint source-channel coding (JSCC). There are various schemes to jointly optimize different components in source and channel coding such as vector quantization and index assignment  \cite{vosoughi2014joint}\cite{heinen2005transactions}\cite{goodman1988using}\cite{bozantizis2000combined}. However, these methods can not provide satisfying performance and most systems still adopt separate coding methods. 

In recent years, deep learning has been successfully applied to a wide range of areas such as computer vision \cite{krizhevsky2012imagenet}\cite{simonyan2014very}, natural language processing \cite{cho2014learning}\cite{vaswani2017attention} and wireless communications \cite{ye2017power}\cite{yang2019ilps}\cite{hsiao2021super}, achieving significant performance improvement over analytical (i.e., non-data-driven) algorithms. Its ability to extract complex features from images has led to applications of deep JSCC over simple channel models such as AWGN and Rayleigh flat fading channels. The authors in \cite{farsad2018deep} first introduce deep learning (DL) to JSCC by using RNNs for text transmission, and in \cite{zarcone2018joint} the authors apply deep JSCC for analog data compression and storage. The first use of deep JSCC for wireless image transmission is reported in \cite{bourtsoulatze2019deep}, where the authors incorporate both source and channel coding into an auto-encoder structure, and source compression and error correction coding are jointly learned through back-propagation optimization. The proposed deep JSCC algorithm in \cite{bourtsoulatze2019deep} outperforms conventional separate schemes under AWGN and Rayleigh flat fading channels. Later, this scheme is extended with larger neural networks \cite{burth2020joint}, feedback \cite{kurka2020deepjscc}, progressive transmission \cite{kurka2019successive} or additional attention modules that deal with different SNRs \cite{xu2020wireless}\cite{ding2021snr}. The idea of deep JSCC is also adopted for discrete channels such as the binary symmetric channel (BSC) \cite{choi2019neural}\cite{song2020infomax}. Due to the discrete nature, the authors perform end-to-end optimization in a variational manner and optimize a lower bound of mutual information between the source images and noisy channel outputs. Different from prior JSCC works, we consider JSCC in the more realistic and challenging multipath fading channels. To the best of  our knowledge, there is no existing JSCC work specifically designed and optimized for the multipath fading channels.

Directly replacing the simple channel model such as AWGN with the multipath fading channel in the prior JSCC works may lead to sub-optimal performance due to the inter-symbol interference. Motivated by model-based deep learning approaches\cite{he2019model}\cite{shlezinger2020model}\cite{o2017introduction}\cite{samuel2019learning}\cite{he2018model}\cite{khani2020adaptive}, we incorporate orthogonal frequency division multiplexing (OFDM) as our domain knowledge, where equalization is performed in frequency domain for orthogonal sub-carriers without inter-symbol interference. Instead of naively treating a neural network as a black box, our model-based deep learning framework imposes a guided structure for neural networks with analytical models and signal processing blocks designed using the expert domain knowledge. The proposed model-based structure shortens the training time and also improves the system performance.

Recently, substantial effort has been made to incorporate end-to-end learning based communication systems with OFDM as domain knowledge. The authors in \cite{felix2018ofdm} first introduce an end-to-end learning-based OFDM system, where an autoencoder maps the transmitted binary bits to baseband channel inputs and recovers the bit information from noisy channel output. This method can be further extended to reduce the high peak-to-average power ratio (PAPR) in OFDM \cite{kim2017novel}. In \cite{balevi2019one}, the authors consider one-bit quantization of the received OFDM signal. A pilot-less end-to-end learning for OFDM is studied in \cite{aoudia2020end}.  However, those prior works only focus on learning modulation constellation designs for arbitrary binary bit sequence inputs without considering the source coding. 
Other prior works such as \cite{ye2017power}\cite{gao2018comnet}\cite{liao2019chanestnet}\cite{van2020optcomnet} limit the scope to deep learning based channel estimation and detection instead of the end-to-end OFDM system.

In this paper, we propose a deep learning based JSCC scheme for wireless image transmission under multipath fading channels. Unlike prior works \cite{bourtsoulatze2019deep}\cite{kurka2020deepjscc}\cite{kurka2019successive} where the encoded images are directly transmitted through the channel, we concatenate deep neural networks with OFDM processing blocks by feeding the neural network encoded image as frequency domain OFDM baseband symbols. The multipath channel and OFDM baseband processing blocks are all represented by non-trainable but differentiable layers, and the whole system can be trained in an end-to-end manner including the multipath channel model.

Numerical experiments show that 
combining JSCC with OFDM in the proposed way outperforms the direct channel transmission used in prior works \cite{bourtsoulatze2019deep}\cite{kurka2020deepjscc}\cite{kurka2019successive}. Moreover, we demonstrate that adopting more domain-knowledge-driven structures such as channel estimation and equalization with optional channel state information (CSI) feedback further improves the system performance. Our results show that the proposed JSCC-OFDM method outperforms the baseline separate source and channel scheme with a state-of-the-art image compression method and error correction code applied to OFDM. Furthermore, enlightened by recent works in image compression \cite{agustsson2019generative}\cite{mentzer2020high}, we demonstrate that our scheme can be combined with adversarial training to produce more visually appealing results. Finally, we show that our proposed method is robust to additional signal clipping-based PAPR reduction\cite{ochiai2002performance}, a wide range of noise power levels, and various channel conditions that the neural network has never seen during training.

Main contributions of this paper are summarized as follows:

\begin{itemize}
  \item To the best of our knowledge, it is the first to extend deep learning based JSCC to  the challenging multipath channels. We jointly train deep neural networks with explicit OFDM blocks to combat the multipath fading and enhance the system performance. 
  \item We design a decoder structure that combines deep neural networks with additional non-trainable blocks such as PAPR reduction signal-clipping, channel estimation and channel equalization to exploit the domain expert knowledge. We conduct detailed ablation studies and verify that this elaborate structure outperforms a blackbox network that solely relies on data-driven training. 
  
  \item The proposed JSCC is evaluated for wireless image transmission compared to a traditional modularized scheme using a state-of-the-art image compression codec and separate error correction code.  It is shown that the proposed JSCC significantly (2.5 - 4dB SNR gain) outperforms the baseline.
  
  \item We evaluate a CSI-feedback based pre-coding scheme to allocate the power for each subcarrier. Our results quantify the gain from the proposed pre-coding that outperforms conventional water-filling algorithms applied to the baseline.
  
  \item We evaluate a new adversarial training scheme for JSCC and demonstrate that it enhances the image perceptual quality. It is also shown that the proposed JSCC method is robust to various channel conditions such as noise power, number of multipaths that the neural networks have not seen during training.
\end{itemize}

\section{Proposed Method}
\figofdmlarge
\subsection{Deep JSCC in multipath fading channels}
The general framework for deep learning based JSCC is shown in Figure \ref{fig:fig_intro}, bottom. In our approach, a pair of encoder $E_\theta$ and decoder $D_\phi$ directly map the real-valued source (image) input $x$ to complex-valued (modulated) baseband samples $y$ and estimate the source signal $\hat{x}$ from noisy multipath fading channel outputs $\hat{y}$. Here, $\theta$ and $\phi$ represent the parameters for the encoder and decoder neural networks respectively. The channel can be represented with the conditional probability $p(\hat{y}|y; \epsilon)$ where $\epsilon$ denotes the parameters that describe the channel statistics. The channel model (\ref{eq:channel}) is defined in the discrete time domain assuming interpolation and Nyquist sampling for the channel inputs and outputs.

We parameterize the multipath fading channel using a discrete channel transfer function:
\begin{equation}
    \hat{y} = h(y;\sigma_0^2, \ldots, \sigma_{L-1}^2, \sigma^2) = h * y + w,
    \label{eq:channel}
\end{equation}
where $*$ denotes the convolution operation, $h \in \mathbb{C}^L$ denotes the sample space channel impulse response, and $L$ is the number of multipaths. $w \sim \mathcal{CN}(0, \sigma^2I_k) $ represents the additive Gaussian noise. Each path experiences independent Rayleigh fading satisfying $h_l \sim \mathcal{CN}(0, \sigma_l^2)$ for $l=0,1,\ldots, L-1$. The power for each path follows an exponential decay profile $\sigma_l^2 = \alpha_l e^{-\frac{l}{\gamma}}$, where $\alpha_l$ is a normalization coefficient to satisfy $\sum_{l=0}^{L-1} \sigma_l^2 = 1$. $\gamma$ is the time decay (or delay spread) constant. Note that the channel transfer function is fully differentiable for $y$ given the realizations of random channel parameters. This means that the gradients from the decoder can propagate back to the encoder through a particular realization of the multipath fading channel $h$ when the end-to-end system is trained with gradient descent methods.

\subsection{JSCC with OFDM extension}

We extend the JSCC framework to an OFDM-based system to efficiently mitigate the multipath fading channel with simple single-tap frequency domain equalization. We assume that each source image $x$ is transmitted in a single OFDM packet that contains $N_p$ pilot symbols and $N_s$ information symbols. For channel estimation, we adopt block-type pilots by sending known symbols on all subcarriers. For simplicity, we assume that the transmitter and receiver are synchronized in time (within the cyclic-prefix length $L_{cp}$) without any carrier frequency offset.  

The OFDM extension to the JSCC framework is shown in Figure \ref{fig:fig_ofdm}. Given the input image dimension of $H \times W$, the encoder output dimension is $C \times \frac{H}{2^d} \times \frac{W}{2^d}$, where $d$ is the down-sampling factor and $C$ is the number of output channels. At this point, the encoder output is real-valued. Then, the encoder output is converted to complex numbers by mapping one half to the real part and the other half to the imaginary part. Then it is reshaped to $Y \in \mathbb{C}^{N_s \times L_{fft}}$. $L_{fft}$ denotes the number of subcarriers in an OFDM symbol and $Y$ represents the frequency domain complex-valued symbols to be fed to the OFDM transmitter. Note that $N_s=\frac{CHW}{2^{2d+1}L_{fft}}$ which is the number of OFDM information symbols. In all of our experiments, the parameters are carefully chosen so that $N_s$ is an integer. The pilot symbols $Y_p \in \mathbb{C}^{N_p \times L_{fft}}$ are known to both the transmitter and receiver.  In this paper, we assume that the $N_p$ pilot symbols are identical without loss of generality. Next, we apply the inverse discrete Fourier transform (IDFT) to each OFDM symbol and append the cyclic-prefix (CP). After that, the transmit signal $y \in \mathbb{C}^{(N_s+N_p)(L_{fft}+L_{cp})}$ propagates through the multipath channel described in (\ref{eq:channel}). Once the receiver obtains the noisy channel output $\hat{y}$, it removes the CP, and applies DFT to produce the frequency domain pilots $\hat{Y}_p$ and data symbols $\hat{Y}$. Thus, the following equations hold for pilot and information symbols:
\begin{equation}
    \hat{Y}_p[i,k] = H[k]Y_p[i,k] + W[i,k],
    \label{eq:h1}
\end{equation}
\vspace{-0.5cm}
\begin{equation}
    \hat{Y}[j,k] = H[k]Y[j,k] + V[j,k],
    \label{eq:h2}
\end{equation}
where $H[k]$ denotes the channel frequency response for the $k_{th}$ subcarrier, and both $W$ and $V$ denote noise samples. The decoder is trained to estimate transmitted source $\hat{x} = D_\phi(Y_p,\hat{Y},\hat{Y}_p)$ given $Y_p$, $\hat{Y}_p$ and $\hat{Y}$. The reconstruction loss can then be expressed as follows:
\begin{equation}
    L_{rec} = \mathbb{E}_{p(x, \hat{Y}, \hat{Y}_p)}\big[||D_\phi(Y_p,\hat{Y},\hat{Y}_p)-x||_2^2\big],
    \label{eq:rec_loss}
\end{equation}
where $\mathbb{E}_p[\ ]$ is the expected value over distribution $p$ and $D_\phi(Y_p,\hat{Y},\hat{Y}_p) (=\hat{x})$. Here, we use mean squared error (MSE) loss as our distortion function. 


It must be noted that including OFDM signal processing in the proposed JSSC framework is possible because IDFT/DFT and CP removal/insertion can be treated as linear layers (i.e., matrix vector multiplications) with fixed parameters in the neural network. During back-propagation, gradients can pass through these (I)DFT/CP linear layers.

\figfeedback
\subsection{Decoder design with domain knowledge}

To estimate the source information, one possible approach is to group $\hat{Y}$, $Y_p$ and $\hat{Y}_p$ and directly feed them into a generator network. This approach relies on the deep neural network to learn the underlying relationships in (\ref{eq:h1}) and (\ref{eq:h2}) to perform the (sub)optimal estimation of the source. However, it treats the entire decoder as a blockbox, and hence it faces the problems of slow convergence and sub-optimal performance as argued in \cite{he2019model} for different applications. As an alternative solution, we propose the decoder structure shown in Figure \ref{fig:fig_ofdm} where explicit signal processing kernels such as channel estimation and equalization are included as domain knowledge. In this work, we adopt the per-channel minimum mean square error (MMSE) channel estimation method used in \cite{felix2018ofdm}, which is as follows:
\vspace{-0.2cm}
\begin{equation}
    H_{MMSE}[k] = \frac{\sum_{i=1}^{N_p}\hat{Y}_p[i,k] Y_p[i,k]^*}{N_p + \sigma^2}. 
    \label{eq:mmse_ce}
\end{equation}

The pilot symbols $Y_p[i,k]$ are designed to have unit power. Here, we use the per-channel MMSE channel estimation for two reasons. First, this method only requires element-wise operations and thus is computationally efficient for the end-to-end training. Second, this method does not require second order statistics of the channel. For equalization, we adopt a conventional MMSE equalizer as shown in (\ref{eq:mmse_eq}). The channel estimation and equalization are included as pre-processing steps in front of the generator network $G$ in the decoder as shown in Figure {\ref{fig:fig_ofdm}}.
\vspace{-0.2cm}
\begin{equation}
    Y_{MMSE}[k] = \frac{Y[i,k] \hat{H}[i,k]^*}{|\hat{H}[i,k]|^2 + \sigma^2}. 
    \label{eq:mmse_eq}
\end{equation}

However, since the explicit channel estimation and equalization may not be optimal, we further introduce two residual connections with two light-weight networks CE subnet ($\Phi_{ce}$) and EQ subnet ($\Phi_{eq}$), so that they can learn and compensate residual errors in the channel estimation and equalization, as shown in Figure \ref{fig:fig_ofdm}. Consequently, we introduce an additional channel estimation loss term to guarantee that the input for equalization is a valid channel estimation. We use a simple MSE loss for the channel estimation loss:
\begin{equation}
\begin{split}
    L_{cha} &= \mathbb{E}\big[||\hat{H}-H||_2^2\big]\\
    &= \mathbb{E}\big[||H_{MMSE}+\Phi_{ce}(H_{MMSE}, Y_p, \hat{Y}_p)-H||_2^2\big].
\end{split}
    \label{eq:ce_loss}
\end{equation}

For the proposed end-to-end system, we jointly optimize the reconstruction loss and channel estimation loss. The total loss can be written as follows:
\begin{equation}
    L_{total} = L_{rec} + \lambda_c L_{cha},
    \label{eq:loss}
\end{equation}
where $\lambda_c$ represents the weight for channel estimation loss. Notice that the back-propagation training of  neural networks requires the partial derivative expression of the channel estimation function (\ref{eq:mmse_ce}) and equalization function (\ref{eq:mmse_eq}), which is straightforward to obtain.

\subsection{OFDM with PAPR reduction} \label{papr_intro}

One notable drawback of OFDM is the high PAPR that causes excessive power consumption at the power amplifier \cite{li1997effects}. For the TX signal $y[n]$, the PAPR is defined as:
\begin{equation}
    \text{PAPR}(y) = \frac{\max_{\substack{n}} |y[n]|^2}{\mathbb{E}|y[n]|^2}.
\label{eq:papr}
\end{equation}
A number of solutions have been proposed to reduce the PAPR \cite{kim2017novel}. In this paper, we investigate the signal clipping technique \cite{ochiai2002performance}\cite{li1997effects}, which can be easily combined with our JSCC scheme.

Signal clipping is one of the simplest and extensively studied PAPR reduction methods. Suppose we have the original TX sample $y[n] = A[n]e^{j\phi[n]}$, where $A[n]$ is the amplitude and $\phi[n]$ is the phase. The clipping operation can be expressed as follows:
\begin{equation}
    y_{clip}[n] = \begin{cases}
A[n]e^{j\phi[n]} &\text{when $A[n] \leq \rho \sqrt{P_s}$}\\
\rho \sqrt{P_s}e^{j\phi[n]} &\text{otherwise},
\end{cases}
\label{eq:clip}
\end{equation}
where $\rho$ denotes the clipping ratio (CR) and $P_s$ is the input signal power. In the proposed JSCC, this clipping operation can be considered as introducing an additional non-linear activation function to the time domain OFDM signal. After clipping, we re-scale (i.e., normalize) the clipped signal so that the signal power is unchanged. According to (\ref{eq:papr}), the normalization operation does not affect the PAPR. Gradient calculation during the training can be back-propagated through those normalization and non-linear clipping functions such that the encoder $E_\theta$ and decoder $D_\phi$ are trained for PAPR reduction via signal clipping.

\subsection{Adaptive power allocation with CSI feedback}

It is known that that although CSI feedback does not increase the capacity of a memoryless channel \cite{Shannon1956}, it can significantly improve the error rate because of the increased error exponent using dynamic power allocation algorithms (e.g., water-filling \cite{cover1991elements}) and adaptive modulation \cite{goldsmith1997variable}. For the system scenario where the CSI knowledge is available, we design an optional (i.e., only enabled when CSI is available) pre-coding block for the proposed JSCC to evaluate the impact of exploiting the CSI knowledge. The structure of our optional pre-coding block is shown in Figure \ref{fig:fig_feedback}. The pre-coding network takes the encoder output $Y$ and (perfect) channel frequency response $H\in\mathbb{C}^{N_s\times L_{fft}}$ as inputs, and generates weights $W\in\mathbb{R}^{N_s\times L_{fft}}$. The weight $W$ mimics the power distribution in traditional power allocation methods and thus it has the same size as $Y$. The OFDM frequency-domain symbols are obtained by doing element-wise multiplication between $W$ and $Y$.

\subsection{Combination with adversarial loss}

Most prior works on JSCC wireless image transmission optimize the traditional distortion metric of PSNR. However, when the image is encoded with a low rate, PSNR tends to lose their importance as it favors pixel-wise error minimization over preservation of texture and content of the image. Thus, aggressive image compression maximizing PSNR could result in blurry images and the obtained PSNR often does not align well with human perception quality \cite{tschannen2018deep}\cite{santurkar2018generative}. Recently, minimizing adversarial loss \cite{goodfellow2014generative} was proposed as a promising solution to capture the global semantic information and local textures, yielding visually-appealing generated images \cite{ledig2017photo}\cite{agustsson2019generative}\cite{mentzer2020high}\cite{liu2020unified}. In this section, we extend our model to an adversarial setting where an additional discriminator neural network is introduced to distinguish whether the image is original (uncompressed) or generated by the decoder (shown in Figure \ref{fig:fig_ofdm}). We adopt the formulation of LSGAN\cite{mao2017least} because of its simple formulation and improved stability than the original GAN \cite{goodfellow2014generative}. The adversarial loss is defined in (\ref{eq:adv}) where $Dis(x)$ is the discriminator network output for input image $x$ (desired output is 1 for uncompressed images and 0 for images generated by the decoder). 
\begin{equation}
    L_{GAN} = \max_{Dis}\mathbb{E}[(Dis(\hat{x})-1)^2] + \mathbb{E}[Dis(x)^2].
    \label{eq:adv}
\end{equation}

When the adversarial loss is adopted, the training process becomes a minmax optimization and the objective function is modified from (\ref{eq:loss}) to (\ref{eq:loss_gan}), where $\lambda_g$ denotes the weight for the adversarial loss.
\begin{equation}
    \min_{E, D} L_{total} = L_{rec} + \lambda_c L_{cha} + \lambda_g L_{GAN}.
    \label{eq:loss_gan}
\end{equation}

\fignet
\tablelow
\tablehigh
\tablecomplex
\section{Training and Evaluation} \label{train}

We first test our proposed method with  images in CIFAR-10, CIFAR-100, and CelebA datasets for wireless communication. CIFAR-10 and CIFAR-100 contain 60,000 32$\times$32-pixel images and we adopt reflection padding and random cropping for data augmentation. CelebA contains more than 200,000 celebrity images. As the pre-processing, we scale and crop CelebA images to 64$\times$64-pixels. For testing, we take the first 10,000 images from each test dataset (unused for training) and transmit each image 5 times through different random realizations of the multipath channel. In section \ref{ablation}, \ref{comparison}, \ref{papr}, and \ref{robust}, we only consider the image reconstruction loss and channel reconstruction loss, and we use both PSNR and SSIM to evaluate the image reconstruction quality. The effect of including an adversarial loss is evaluated in section \ref{adversarial}. Despite the small image size described here, our JSCC scheme can also be applied to high-resolution images. More details are introduced in section \ref{large}. 

The encoder and decoder networks for CIFAR-10 and CIFAR-100, which follow the design principles in \cite{isola2017image}, are shown in Figure \ref{fig:fig_net}. For the encoder, each image is first normalized within the range of $[-1,1]$. Then, it is passed through a series of down-scaling convolutional layers and residual blocks. The output of the final layer consists of $C$ channels and it is followed by a power normalization layer to enforce unit power. Note that we can change $C$ to control the compression rate. The structure of the generator network is almost symmetric (in the reverse order) to the encoder network, except that the final activation function is set to Sigmoid function to enforce a valid dynamic range of image output pixels. Networks for CelebA have one more down-sampling module at the encoder and one more up-sampling module at the generator network. The structure of the CE subnet and EQ subnet is kept the same for all datasets. Here, we use convolutional layers instead of fully connected layers because of their efficiency in terms of network size (number of parameters). Although the input signals for the CE subnet and EQ subnet are 1D in nature, we reshape them to be 2D so that we can use a small kernel to obtain the full receptive field covering the entire signal. For example, assume the 3 inputs (i.e., $Y_p, \hat{Y}_p$, and $H_{MMSE}$) for the CE subnet have a size of $2\times64$. We could reshape and stack them to form an input of $6\times8\times8$. Then, three 2D convolutional layers with a kernel size of $5\times5$ are sufficient to guarantee a full receptive field. The same logic applies to the EQ subnet. 

We optimize the loss function in equation (\ref{eq:loss}) and set $\lambda_c$ to $0.5$ for all experiments in this subsection. We use a batch size of 128 for CIFAR-10 and CIFAR-100 and a batch size of 64 for CelebA. All neural networks are trained using ADAM with $\beta_1=0.5$, $\beta_2 = 0.999$, and an initial learning rate of $5\times10^{-4}$. We train for 300 epochs and 30 epochs for CIFAR-10/CIFAR-100 and CelebA respectively. For all datasets, we apply linear learning rate decaying for the second half of the training process.    

As for the OFDM system, the parameters are set to $L_{fft}=64, L_{cp}=16, L=8$, and $\gamma=4$. Note that the coding rate of our method depends on the number of samples after encoding. For a source image of the size $H\times W \times C$ pixels ($C$ is the number of color channels), the coding rate is $\frac{(N_p+N_s)(L_{fft}+L_{cp})}{HWC}$ in channel-use per pixel (CPP). The length of CP has to be larger than the delay spread of the channel to avoid inter-symbol interference. 

Note that all evaluations in the following subsections except for Section \ref{feedback} are performed without the CSI feedback (i.e., the pre-coding network in Figure \ref{fig:fig_feedback} is unused), and we use instantaneous SNR defined at the receiver, which is:
\begin{equation}
    \text{SNR} = 10\log_{10}\frac{\mathbb{E}|\hat{y}|^2}{\sigma^2} (\text{dB}).
\label{eq:snr}
\end{equation}

\subsection{Effect of domain knowledge and residual connections} \label{ablation}
\figresult
\figresultlarge
In this section, we quantify the effect of incorporating different amounts of domain knowledge into our framework and the effect of additional subnets. For each method, we evaluate the PSNR, SSIM and channle estimation mean square error (CE MSE) if available. In Table \ref{tab1} and \ref{tab2}, the first row  represents a simple JSCC scheme where symbols are transmitted directly through the multipath channel without any explicit OFDM layers (as in Figure \ref{fig:fig_intro}, bottom). In the second row, we introduce the OFDM blocks (DFT, CP, etc.) as our domain knowledge. However, for the decoder network, we directly concatenate $Y_p, \hat{Y_p}$ and $\hat{Y}$ and feed them to a generator network without any explicit channel estimation and equalization. Next, in the third row, we further introduce per-channel MMSE channel estimation and MMSE equalization as extra domain knowledge. However, we do not include any residual connections for refinement. Next, for the fourth row, we add CE subnet and the channel reconstruction loss to refine the channel estimation. Finally, in the fifth row, we evaluate the proposed system with the second residual connection and EQ subnet. All of the above methods are tested with CIFAR10, CIFAR-100 and CelebA under two different SNRs (5dB and 15dB). For the last four rows, where the OFDM system is included, we set $N_s=7$ and $N_p=1$, which provides a CPP rate of 0.208 for CIFAR-10/CIFAR-100 and a CPP rate of 0.052 for CelebA. For the direct transmission scheme, we keep the same CPP by properly sizing the encoder output dimension for a fair comparison.

From row 1 and row 2, we can see that when we introduce OFDM processing blocks to alleviate the multipath fading, the performance is greatly improved for both SNRs and across all datasets, especially for the 15dB SNR case where there is over 3dB gain in PSNR and 0.1 gain in SSIM for all tested datasets. When we further add explicit channel estimation and equalization, we observe a small improvement in the high SNR case. However, for the 5dB SNR case, both PSNR and SSIM drop instead. We think the main reason is that, when SNR is low, the error of per-channel MMSE channel estimation is so large that the benefits derived from domain knowledge is cancelled out. Whereas when the SNR is high, the channel estimation error is relatively small, and therefore the benefit from domain knowledge is evident. As we add CE subnet and the channel estimation loss to refine our channel estimation, we observe better channel estimation quality and we finally get better reconstruction quality for both 5dB and 15dB SNR. Lastly, as we further add EQ subnet and complete our final proposed scheme, we get the best performance among all the variants.

We provide the complexity analysis for the networks used for CIFAR-10 as shown in Table \ref{tab3}. OFDM processing complexity compared to the generator network is insignificant and the all subnets are much smaller than the generator network. Hence all methods listed in Table \ref{tab3} have comparable complexity. In fact, slight complexity reduction from 'Direct' method to OFDM-based methods (OFDM+CE+EQ) is achieved as OFDM pilot symbols do not require any neural network processing. Compared to 'OFDM+CE+EQ', the CE and EQ subnet in our proposed method (Figure \ref{fig:fig_ofdm}) only takes $\approx 1.2\%$ additional parameters to achieve significant performance enhancement.

\subsection{Comparison with a separate coding scheme} \label{comparison}
\figresultsnr
In this section, we compare our proposed JSCC method with a baseline separate source and channel coding scheme. The baseline separate coding scheme uses the state-of-the-art BPG \footnote[1]{https://bellard.org/bpg/} as the image codec and LDPC in the IEEE 802.11n WiFi standard \footnote[2]{https://github.com/tavildar/LDPC} as the channel code. We test with three different LDPC codes, $(972, 1944), (1296, 1944), (1458, 1944)$, which correspond to rates $1/2$, $2/3$ and $3/4$, respectively. We use BPSK, QPSK, 16QAM, and 64QAM for OFDM modulation applied to BPG- and LDPC-coded bit sequences. Then, we enumerate all possible modulation and coding rate combinations to identify the optimal configuration for the baseline separate coding scheme. We consider both perfect channel state information (CSI) and the per-channel MMSE channel estimation used in our proposed scheme. The estimated channel is then used to calculate the log likelihood ratio (LLR) for LDPC decoding in the baseline separate coding scheme. Note that the perfect CSI gives an (unfair) advantage to the baseline method and serves as an upper bound for all channel estimation methods.

We compare PSNR and SSIM of the proposed JSSC and the separate coding baseline at the same/similar rate and SNR. Because BPG cannot achieve arbitrary compression rates, we exhaustively search for the BPG parameter to obtain a rate closest to the target. For fair comparison, we remove the BPG headers in the BPG rate computation. When the decoded BPG bit sequence contains bit errors, either 1) it is decompressed to an image with large quality degradation or 2) it fails to decode any valid image due to catastrophic failures in the decoder algorithm. When the latter happens, we assume (although it does not produce any valid image) that the corrupted BPG bit sequence corresponds to the average PSNR/SSIM of images recovered in the aforementioned case 1.

Figure \ref{fig:fig_result1} compares the performance of deep JSCC schemes and baseline approaches with respect to the number of information symbols $N_s$ in different SNR regimes using the CIFAR-10 dataset. A smaller $N_s$ indicates a more aggressive (lower) rate to send an entire image with a lower number of data symbols. For comparison, we show the performance of baseline separate coding schemes with 1) perfect CSI for error-free channel estimation, and 2) per-channel MMSE channel estimation with different numbers of pilots. We only show the performance of the best combination of source and channel coding, which is BPG+1/2LDPC+QPSK for 5dB and BPG+1/2LDPC+64QAM for 15dB.

From Figure \ref{fig:fig_psnr1}, \ref{fig:fig_ssim1}, we can see that the proposed JSCC scheme consistently outperforms the baseline separate coding scheme with perfect CSI, especially when $N_s$ is small. When the channel estimation accuracy drops due to fewer number of pilots ($N_p$) for per-channel MMSE estimation, our JSCC method outperforms the baseline method by a larger margin although it uses only $N_p=1$ pilot symbol. Even when provided with perfect CSI, the separated coding schemes could not outperform the JSCC method. In practice, the baseline scheme may need a substantial number of pilot symbols to approach near-optimal channel estimation accuracy, which can significantly decrease the transmission rate. As shown in Figure \ref{fig:fig_psnr2}, \ref{fig:fig_ssim2}, when tested with 15dB SNR, the baseline separate coding scheme with perfect CSI approaches the performance of the proposed JSCC method in PSNR when $N_s$ is large. This observation aligns with prior works \cite{bourtsoulatze2019deep}\cite{kurka2020deepjscc} and shows that the JSCC method achieves the biggest advantages compared with separated coding schemes in the low SNR and low rate region. Further, it is observed that our JSCC method consistently outperforms the baselines although channel estimation is non-optimal with only $N_p=1$, especially in the low SNR condition. Also, our JSCC method is consistently better than all baselines in SSIM that is better aligned with human perception than PSNR.

Figure \ref{fig:fig_mse1} and \ref{fig:fig_mse2} show the trade-off between the pilot symbols and information symbols given a fixed total transmission symbol budget ($N_s+N_p=7$, $\text{CPP}=0.182$) for the constant packet length. On one hand, we prefer more pilot symbols so that we can achieve better channel estimation accuracy. On the other hand, it is beneficial to use more information symbols to carry the transmitted images with more redundancy in coding. Our results with both 5dB and 15dB SNRs show that as JSCC uses more pilot symbols, both the channel estimation MSE and PSNR improve. It implies that using only 1 pilot symbol for channel estimation ($N_p=1$) and remaining symbols for data $N_s=6$ is optimal for the JSCC method as shown in Figure \ref{fig:fig_mse1} and \ref{fig:fig_mse2}. Although less pilots tend to decrease channel estimation accuracy, the overall image reconstruction quality improves by the gain of using more data symbols to encode the transmitted image.

Next, in Figure \ref{fig:fig_result3}, we fix the total transmission budget to $N_s+N_p=7$ ($\text{CPP}=0.182$) and evaluate the image reconstruction quality for a variety of SNRs. For the baseline method, we choose at each point the best combination of the OFDM modulation QAM size and LDPC rate. Since we need at least one pilot symbol for channel estimation, the curve for $N_p=1$ and perfect CSI is essentially an upper bound of all the baselines considered in this paper. Our proposed method consistently outperforms the baseline method's upper bound for both PSNR and SSIM across all SNRs. When we increase the number of pilot symbols (e.g., $N_P=3$) for better CSI estimation in the baseline method, the performance gap increases between the baseline and the proposed JSCC. Furthermore, for the optimal case of $N_p=1$, the proposed JSCC outperforms baseline methods more significantly when SNR is lower.

\subsection{Performance with PAPR reduction} \label{papr}
\figresultpapr
\figresultclip
\figfeedbackresult
In this section, we evaluate our proposed method with additional signal clipping as our PAPR reduction technique. We first compute the average PAPR and PSNR obtained by different clipping ratios and the results are shown in Figure \ref{fig:fig_papr}. We train our model with four clipping ratios $\rho=0.8, 1, 1.2, 1.4$ for both SNR$=$5dB (blue markers) and SNR$=$15dB (red markers). For all points in that figure, we set $N_s=6$, $N_p=1$ ($\text{CPP}=0.182$). To show the effect of end-to-end training with clipping, we compare it (solid line) with a simple scenario where the clipping operation is applied after training neural networks without exposing them to clipping during the training (dash line).   

Figure \ref{fig:fig_papr} demonstrates the trade-off between image reconstruction quality and PAPR. When no PAPR reduction is applied, ($\rho=\infty$), we obtain the highest PSNR but also the highest PAPR around 10dB for both noise levels, which puts a significant amplifier power efficiency burden for practical systems \cite{ochiai2002performance}. As we gradually put constraints on the PAPR (smaller $\rho$), we obtain lower PAPR as desired, but we gradually lose the representation power of our model and the image quality drops as well. With carefully designed clip ratios, we are able to achieve a low PAPR under 2dB for both 5dB and 15dB SNR. We can also observe that our model is more resilient to the clipping noise in 5dB than 15dB. When the SNR is 5dB, the channel noise dominates the total impairment; thus, the clipping noise has little effect on the image quality. However, when the SNR is 15dB, the clipping noise dominates the total distortion and hence has a large impact on the quality drop. When we directly apply the clipping operation without training with it, the additional non-linear noise has never been seen by the neural networks and thus leads to a larger performance degradation, which is shown by the dash lines in Figure. \ref{fig:fig_papr}. Similarly, the clipping noise has a larger impact for the high SNR case.

Figure \ref{fig:fig_clip} compares the effect of clipping on the performance of the proposed JSCC and baseline method with various clipping ratios (CR) of $\rho=1$, $1.4$, and $\infty$ as defined in (\ref{eq:clip}). Note that for the baseline, we adopt the methods in \cite{ochiai2002performance}\cite{kim2011power} to calculate the LLR including the non-linear distortion introduced by clipping. It is  observed that our JSCC approach provides a more gradual degradation in PSNR/SSIM than is observed for the baseline method as more aggressive clipping is applied. Similar to that reported in the previous section, our method outperforms the baseline approach for all SNRs tested in the presence of clipping. 

\subsection{Performance with CSI feedback} \label{feedback}

In this section, we evaluate the performance of the proposed system when the CSI feedback is available. We assume the perfect CSI is available at both transmitter and receiver. Hence, we remove the channel estimation module and CE subnet shown in Figure \ref{fig:fig_ofdm} and activate the optional pre-coding network shown in Figure \ref{fig:fig_feedback}. The structure of the pre-coding network follows the same structure as the subnet in Figure \ref{fig:fig_net}. We compare our JSCC method with/without CSI feedback to a baseline separated scheme that uses a water-filling power allocation algorithm \cite{cover1991elements}, which assigns the optimal power for each subcarrier to attain the channel capacity when the CSI feedback is available.  Instead of devising an optimal modulation and channel coding scheme (which is non-trivial) for the baseline, we make a simplifying assumption that it achieves the channel capacity given CSI. It corresponds to an upper bound performance for the baseline using BPG for image coding. The results are shown in Figure  \ref{fig:fig_feedback_result} where $N_s=6$. Notice that the instantaneous (receiver) SNR defined in (\ref{eq:snr}) is no longer compatible when the transmiter performs the water-filling power allocation based on the CSI feedback. Thus, when we evaluate the CSI feedback performance, we adopt the frequency domain average SNR defined as:
\begin{equation}
\begin{split}
    \text{Average SNR} &= 10\log_{10}\frac{\frac{1}{L_{fft}}\sum_{l=0}^{L_{fft}-1}\mathbb{E}|H[l]X[l]|^2}{\sigma^2}\\
    &= 10\log_{10}\frac{\mathbb{E}|H|^2 P}{\sigma^2} = 10\log_{10}\frac{P\sum_{l=0}^L \sigma^2_{l-1}}{\sigma^2},
\end{split}
\label{eq:averagesnr}
\end{equation}
where $P$ is the signal power in frequency domain. As can be seen in Figure \ref{fig:fig_feedback_result}, exploiting the feedback of perfect CSI improves the performance for both PSNR and SSIM. Compared with the upper bound of the baseline method that fully attains the channel capacity, our proposed JSCC method provides similar PSNR and consistently better SSIM. The gain from the CSI feedback is larger for the proposed JSCC than that of the baseline separate scheme.

\subsection{Performance with adversarial loss} \label{adversarial}

In this section, we train our JSCC method with the adversarial loss (\ref{eq:loss_gan}) and we call this method JSCC-\textit{adv}. To better evaluate the perceptual quality or recovered images, we adopt an additional metric called classification accuracy (CA) that evaluate the classification performance of the generated images based on a network pre-trained on original images. For CIFAR-10, we pre-train a ResNet50 network \cite{he2016identity} as the classifier and it achieves a $93.62\%$ classification accuracy on the original CIFAR-10 test set. Some numerical results on CIFAR-10 is shown in Table \ref{tab_adv}. For all compared methods, we set $N_s=6$ and $N_p=1$ ($\text{CPP}=0.182$). The baseline method uses 1/2LDPC+QPSK for 5dB SNR and 1/2LDPC+64QAM for 15dB SNR, and the perfect CSI is assumed to be known. For the JSCC-\textit{adv} method that combines the adversarial loss as in (\ref{eq:adv}), we set $\lambda_g=2\times 10^{-4}$ for 5dB SNR and $\lambda_g=5 \times 10^{-5}$ for 15dB SNR. For both SNRs, the original JSCC method achieves the highest PSNR and SSIM whereas the JSCC-\textit{adv} method achieves the highest CA. That means the images recovered by JSCC-\textit{adv} tends to better align with the distribution of real images although the additional adversarial loss term does slightly degrades the MSE objective thus sacrifices PSNR/SSIM accordingly. This perception-distortion trad-off aligns with prior works such as \cite{blau2018perception}. Although JSCC-\textit{adv} appears lower in traditional distortion measures PSNR and SSIM (due to the adversarial loss term in addition to MSE), it still outperforms the baseline method. It can also be noticed that the perceptual improvement is more obvious for low SNR scenarios. 

Figure \ref{fig:fig_visual} shows visualization examples of recovered images for both CIFAR-10 and CelebA. It can be observed that the JSCC method provides much better recovered images than the baseline method. However, the recovered images are blurry due to optimization of the MSE loss. On the contrary, combined with adversarial loss, JSCC-\text{adv} produces images with more high-frequency details such as human hair and background. This improvement of human perception aligns with the improvement of CA in Table \ref{tab_adv}.

\subsection{Robustness analysis} \label{robust}

\tableadv
\figvisual

\figrobsnr
\figrob
In this section, we evaluate the impact of mismatched channel statistics such as SNR or the number of multipaths used during the JSCC training vs. actual testing. Figure \ref{fig:fig_rob} shows the JSCC performance when it is trained for a specific SNR (5dB or 15dB) and tested with a wide range of SNR from 0dB to 20dB. The model trained with 5dB performs better in the low SNR regime, whereas the model trained with 15dB performs better in the high SNR region. However, both of them still outperform the baseline (with perfect CSI) for most cases. The same trend can be observed from the channel estimation error on the right. Notice that even when the model is tested with unseen noise levels, its channel estimation consistently benefits from the additional residual connection and outperforms the basic per-channel MMSE method. 

We also test the JSCC model with mismatched multipath number ($L$). The JSCC method in Figure \ref{fig:fig_rob} is trained with an $L=8$ multipath channel model and tested with a wide range of $L$ but the same noise level. The system performance is hardly affected with less than 8 paths for both SSIM and channel estimation error. Then the channel has more than 8 paths, the channel becomes more difficult and degradation on both channel estimation and SSIM gradually appears. However, within a certain range of $L$, the JSCC method is robust to channel mismatch and still outperforms the baseline separated methods with perfect CSI.

\subsection{Extension to larger images and qualitative results} \label{large}

\figlarge
In this section, we show that our JSCC scheme can be generalized to high-resolution images. We train our model with 200,000 natural images from Open Image Dataset V6 \footnote[3]{https://storage.googleapis.com/openimages/web/index.html}. Similar with prior JSCC works\cite{bourtsoulatze2019deep}\cite{kurka2020deepjscc}, we train on 256$\times$256 random patches with a batch size of 16. Compared with the network structure in Figure \ref{fig:fig_net}, this model contains one more down-scale convolutional layer and two more residual blocks for the encoder and one more upscale convolutional layer and two more residual blocks for the generator network. Same with the networks for small images, we train our model using ADAM with $\beta_1=0.5$, $\beta_2 = 0.999$ and an initial learning rate of $5\times10^{-4}$. We train our model for 40 epochs and apply linear learning rate decaying for the last 20 epochs.  

Thanks to the fully convolutional structure applied to a patch of large images, we can test our model on multiple input sizes not limited to 256$\times$256. We evaluate our model with the Kodak dataset, which consists of 24 768$\times$512 high-resolution images and serves as a popular benchmark for image compression. In Figure \ref{fig:fig_kodak_result}, we show the comparison between our JSCC and conventional schemes in PSNR and MS-SSIM\cite{wang2003multiscale}. For our JSCC scheme, we set $N_s = 576$ and $N_p=2$, which corresponds to a CPP rate of $0.039$. We transmit each tested image 10 times through random channel realizations. 
The baseline method uses 1/2 rate LDPC and QPSK for the same CPP of 0.039.

As can be seen from Figure \ref{fig:fig_kodak_result}, our method provides similar PSNR in the low SNR region but worse PSNR in the high SNR region. However, when evaluated with MS-SSIM, our method greatly outperforms the baseline scheme. It is worth noting that PSNR is a less reliable (subjective) quality measure for high resolution images in Kodak dataset. For visualization, we present some examples from the 5dB SNR case in Figure \ref{fig:fig_result_kodak}. As shown in Figure \ref{fig:fig_kodak_result}, the images generated by the JSCC method have noticeably better visual quality as indicated by the (MS-)SSIM metric (even when the measured PSNR is lower). When we adopt the adversarial loss (i.e., JSCC-adv), the visual quality is further improved and the reconstructed images successfully recover fine details such as the bush, grass, and ground texture although the obtained PSNR, SSIM and MS-SSIM metrics are lower. Our experiment shows that JSCC can be successfully trained with the adversarial loss for large images and the adversarial loss aligns better with human perception.

\figresultkodak

\section{Conclusion}
In summary, to the best of our knowledge, we present the first implementation of an OFDM-assisted deep JSCC method for multipath fading channels. We combine the deep JSCC scheme with explicit OFDM layers and carefully design the decoder using both expert domain knowledge (such as channel estimation and equalization) and refinement neural networks. Our results demonstrate that such usage of domain knowledge such as OFDM, explicit channel estimation and equalization greatly boost the performance of our model compared to unstructured black box networks. Through extensive experimental simulations, we show that our deep JSCC method outperforms the combination of state-of-the-art image compression codecs, conventional high performing channel codes and OFDM systems, even with perfect CSI. The margin is larger for low SNR and low rate regimes. 
When combined with adversarial loss, our method tends to generate more realistic images. Moreover, the JSCC framework incorporates deliberate signal clipping during the training process to significantly reduce the PAPR with gradual performance degradation. Furthermore, our approach is shown to be robust when the evaluation channel model characteristics are mismatched to those used during the training.

One possible direction of future works is to generalize this work to other types of sources such as audio and video. Different types of sources may require specific model architectures and training strategies to achieve the maximum performance. Another future direction is to add more details to the OFDM system such as carrier frequency offset (CFO) and packet detection so that the system is more practical. One could also design novel pilot patterns to achieve the best efficiency for both channel estimation and source reconstruction.   

\appendices
\section{Ablation Studies on $\lambda_c$ and $\lambda_g$}

An ablation study of $\lambda_c$ and $\lambda_g$ was conducted on CIFAR-10 to better illustrate their effect. For the ablation study of $\lambda_c$, we set $\lambda_g=0, N_s=6, N_p=1$ and $\text{SNR}=5dB$. The results of PSNR and channel estimation MSE are shown in Table. \ref{tablambdac}. Obtained PSNR is not sensitive to $\lambda_c$ although setting $\lambda_c = 0$ results in significantly larger channel estimation error. Using channel estimation subnet $\Phi_{ce}$ enhances the quality of the image as shown in Table \ref{tablambdac}. Although the image quality does not significantly depend on $\lambda_c$ as long as $\Phi_{ce}$ is included in the structure, applying $\lambda_c > 0$ produces more interpretable trained results from the channel estimation subnet.

For the ablation study of $\lambda_g$, we set $\lambda_c=0.5, N_s=6, N_p=1$ and $\text{SNR}=5dB$. As we gradually increase $\lambda_g$, the reconstruction PSNR drops while the classification accuracy increases. When $\lambda_g$ gets too large, however, the neural network tends to create (fake) contents that are differ from original images, degrading both PSNR and classification accuracy. To visualize it, we show examples from CIFAR-10 in Figure \ref{fig:fig_visual_lambdag}.
\tableabc
\tableabg
\figvisuallambda

\section{Comparison with other JSCC structures}
We compare our basic neural network structure without OFDM and subnets (i.e., `Direct w/o OFDM' in Table \ref{tab1} and \ref{tab2}) with those proposed in \cite{bourtsoulatze2019deep} and \cite{burth2020joint}. We conduct experiments on CIFAR-10 using AWGN and multipath fading channels. As we exclude OFDM from our scheme, the difference only comes from the neural network structure itself. For each 32 $\times$ 32 image, we assign 384 channel usages, yielding a CPP of 0.375. The evaluation results are shown in Figure \ref{fig:fig_net_result}. Our structure (even without OFDM and proposed subnets) constantly outperforms the other structures in both AWGN channel and multipath fading channels whereas our structure uses 23\% fewer parameters than the structure in \cite{burth2020joint}. Our performance further improves in the multipath fading channel as we incorporate OFDM blocks as well as proposed subnet structures (`OFDM+CE+EQ+$\Phi_{ce}$+$\Phi_{eq}$') as quantified in Table \ref{tab1} and \ref{tab2}.
\fignetresult


%






\ifCLASSOPTIONcaptionsoff
  \newpage
\fi

\bibliographystyle{IEEEtran}
\bibliography{refs}

\end{document}